\documentclass[review]{elsarticle}
\usepackage[english]{babel}
\usepackage{amsmath}
\usepackage{txfonts}
\usepackage{lineno,hyperref}
\usepackage{setspace}
\usepackage{multirow}
\usepackage{array}
\usepackage{color}
\journal{}
\biboptions{authoryear}
\begin{document}
\begin{frontmatter}
\title{Constraints on the composition and temperature of LLSVPs from seismic properties of lower mantle minerals}
\author[HU,IES]{Kenny Vilella\corref{correspondingauthor}}
\cortext[correspondingauthor]{Corresponding author}
\ead{kennyvilella@gmail.com}
\author[ENS]{Thomas Bodin}
\author[EPFL]{Charles-Edouard Boukar\'e}
\author[IES]{Fr\'ed\'eric Deschamps}
\author[IPG]{James Badro}
\author[UCL,ETH]{Maxim D. Ballmer}
\author[CAS]{Yang Li}
\address[HU]{JSPS International Research Fellow, Hokkaido University, Japan}
\address[IES]{Institute of Earth Sciences, Academia Sinica, Taipei, Taiwan}
\address[ENS]{Univ Lyon, Univ Lyon 1, ENSL, CNRS, LGL-TPE, F-69622, Villeurbanne, France}
\address[EPFL]{Earth and Planetary Science Laboratory, \'Ecole Polytechnique F\'ed\'erale de Lausanne, Lausanne, Switzerland}
\address[IPG]{Institut de Physique du Globe,  Univ. Paris Diderot, Sorbonne Paris Cit\'e, CNRS, Paris, France}
\address[ETH]{Institute of Geophysics, Department of Earth Sciences, ETH Zurich, Zurich, Switzerland}
\address[UCL]{University College London, Department of Earth Sciences, London, UK}
\address[CAS]{State Key Laboratory of Lithospheric Evolution, Institute of Geology and Geophysics, Chinese Academy of Sciences}
\begin{abstract}
Seismic observations have suggested the presence of two Large Low Shear Velocity Provinces (LLSVPs) in the lowermost mantle whose nature and origin are still debated.
Several studies have tried to infer their potential composition using seismic observations with the hope to identify their formation mechanism. 
In particular, compositions enriched in iron ($\sim$~12 -- 14~wt\%) and bridgmanite ($\sim$~90~vol\%) have been identified as promising candidates.
Interestingly, these characteristics are somewhat consistent with the cumulates produced by the solidification of a primitive magma ocean, except that the iron enrichment should be much larger ($>20$~wt\%).
Here, we provide a reappraisal of potential LLSVPs compositions based on an improved mineralogical model including, in particular, the effects of alumina.
We systematically investigate the effects of six parameters: FeO and Al$_{2}$O$_{3}$ content, proportion of CaSiO$_{3}$ and bridgmanite (so that the proportion of ferropericlase is implicitly investigated), Fe$^{3+}$/$\sum$Fe and temperature contrast between far-field mantle and LLSVPs.
From the 81 millions cases studied, only 79000 cases explain the seismic observations.
Nevertheless, these successful cases involve a large range of parameters with, for instance, FeO content  between 12--25~wt\% and Al$_{2}$O$_{3}$ content between 3--17~wt\%.
We then apply a principal component analysis (PCA) to these cases and find two robust results:
(i) the proportion of ferropericlase should be low ($<$6vol\%);
(ii) the formation of Fe$^{3+}$-bearing bridgmanite is much more favored compared to other iron-bearing components.
Following these results, we identify two end-member compositions: a Bm-rich and a CaPv-rich one.
For each end-member composition, a large range of parameters is possible.
We note, however, that a low temperature contrast ($<$500~K) is favored, and that a certain proportion between FeO content, Al$_{2}$O$_{3}$ and oxidation state should be maintained.
Finally, we discuss different scenarios for the formation of LLSVPs and propose that investigating the mineral proportion produced by each scenario is the best way to evaluate their relevance.
For instance, the solidification of a primitive magma ocean may produce FeO and Al$_{2}$O$_{3}$ contents similar to those suggested by our analysis.
However, the mineral proportion of such reservoirs is not well-constrained and may contain a larger proportion of ferropericlase than what is allowed by our results.
%Owing to the lack of available constraints, we were not able to select a preferred scenario or composition.
%From a broader perspective, we conclude that investigating the mineral proportion produced by the different scenarios is probably the best way to evaluate the most likely formation mechanism of LLSVPs.
\end{abstract}
\end{frontmatter}
\section{Introduction}
The emergence of tomographic models has led to an increased knowledge of the structure and composition of the Earth's mantle.
A key result provided by these models is the existence of two large anomalous volumes located at the bottom of the mantle beneath Africa and the Pacific \citep[first reported by][]{Su1994}.
These two nearly antipodal reservoirs, usually referred to as Large Low Shear Velocity Provinces (LLSVPs), cover up to 30\% of the core-mantle boundary (CMB) and may reach a height up to 1000~km \citep[see review by][for more details]{Garnero2016}.
LLSVPs are characterized by a reduction of both the S-wave ($V_{S}$) and P-wave ($V_{P}$) velocities by more than 2\% and 0.5\%, respectively, but an increase in the bulk sound velocity $V_{\phi}$ around 0.5-1\%.
This causes an anti-correlation between $V_{S}$- and $V_{\phi}$-anomalies \citep{Masters2000}, also observed in normal mode data \citep{Trampert2004}.
Early studies have noted that, while a temperature excess may explain the reductions in $V_{S}$ and $V_{P}$, it should also induce a reduction of $V_{\phi}$ that is not observed.
The anti-correlation between the behavior of $V_{S}$ and $V_{\phi}$ was therefore interpreted as an evidence of the thermo-chemical nature of LLSVPs \citep[][]{Ishii1999}, i.e., these structures differ in temperature and composition from the far-field mantle.

Since their discovery, important efforts have been made to determine the nature and origin of LLSVPs.
In particular, several geodynamical studies have explored various possible scenarios that may lead to the formation of structures mimicking LLSVP \citep[][]{Tackley1998b, McNamara2005, Nakagawa2010b, Li2014}.
A common conclusion of these studies is that a reservoir of material denser than the bulk mantle can explain the shape of LLSVPs.
However, it is difficult to estimate precisely the required density excess because other parameters (e.g., rheology) also influence the thermo-chemical structure \citep{Deschamps2008}.
In these works, the authors mainly used the shape of the LLSVPs to assess the relevance of their models neglecting the constraints given by the seismic wave speeds anomalies. 
Alternatively, some studies have used available data from mineral physics to constrain the potential composition of LLSVPs \citep{Samuel2005, Deschamps2012b}.
More specifically, they calculated the density and seismic wave speeds for a large range of compositions and compared them to seismological and geodynamical constraints.
They found that a reservoir enriched in bridgmanite ($\sim$~90~vol\%) and iron ($\sim$~12 -- 14~wt\%), with respect to a pyrolitic composition, provides a good fit to observations.

%Different scenarios have been proposed to explain the formation and existence of LLSVPs.
%The accumulation of subducted materials is a possible way to build such reservoirs \citep[][]{Nakagawa2010b}, although its ability to explain seismic observations remains unclear \citep{Deschamps2012b}.
%Another possibility is the creation of dense reservoirs from the solidification of a primitive magma ocean \citep{Labrosse2007, Ballmer2017}.
%While the general characteristics of this primitive material are somewhat consistent with the inferred composition of LLSVPs \citep{Samuel2005, Deschamps2012b, Boukare2015}, i.e., enriched in bridgmanite and iron, the predicted iron enrichment differs importantly ($>20$~wt\% for the solidification of a magma ocean compared to $\sim$12 to 14~wt\% deduced from seismic data).
%Such a large iron enrichment is believed to induce too large density anomalies to produce reservoirs with the shape and volume of LLSVPs \citep{Ballmer2017}.

Previous studies, however, have left behind several important parameters such as the effects of alumina, ferric-ferrous iron ratios, and spin state transition. 
Here, we use a model including these effects, together with an updated database for the properties of minerals, to investigate the ability of about 81 millions thermo-chemical models to explain the seismic signature of LLSVPs.
Our mineralogical model is characterized by six free parameters that are allowed to vary in large ranges: phase proportions in the mineralogical assemblage (bridgmanite, Ca-silicate perovskite, and ferropericlase, which is dependent on the previous two as their sum must be 100\%), iron and aluminium content in that assemblage, ferric/ferrous iron ratio in bridgmanite (the only phase hosting iron in those two valence states), and temperature.
For each set of parameters, the density and seismic wave speed anomalies are calculated using the Mie-Gr\"uneisen-Debye equation of state.
The incorporation of alumina in bridgmanite changes its bulk modulus, density and the Fe-Mg exchange coefficient between bridgmanite and ferropericlase \citep{Catalli2011, Piet2016, Shukla2016}, while spin state transition affects mainly the density and bulk modulus of ferropericlase \citep{Fei2007, Catalli2010}.
As a consequence, these new features induce important modifications on the calculated properties.
For instance, a much higher content of FeO can be incorporated (up to 25wt\%) without leading to excessively high density.

The identification of plausible thermo-chemical models requires to compare our results with observations.
Generally, available seismic observations provide robust constraints on the seismic wave speed anomalies of LLSVPs.
We therefore calculate these anomalies by comparing the seismic wave speeds obtained for the different thermo-chemical models considered, with the ones obtained for a typical composition and pressure-temperature conditions of the bulk lower mantle. 
The calculated anomalies are then compared to the observed ones.
Due to uncertainties, we consider as successful all the cases with: (i) $V_{S}$-anomalies between -2\% and -7\%; (ii) $V_{P}$-anomalies lower than -0.5\%, both being based on seismic tomography observations \citep[e.g.,][]{Houser2008, Koelemeijer2016}; (iii) positive $V_{\phi}$ anomalies, for example observed in normal mode data \citep{Trampert2004}; (iv) compositional density ($\rho$) increase between 2\% and 2.8\%, based on the values required by geodynamical studies to produce reservoirs with the shape of LLSVPs \citep{Li2014}.
From all the models we tested, a collection of 79000 cases, involving a large range of properties, satisfies these constraints.
We then describe and discuss in details these successful compositions.
\section{Mineralogical model of the lower mantle}
In order to calculate the density and seismic wave speed anomalies of LLSVPs, it is first necessary to know the properties of the bulk (far-field) mantle.
In this section, we provide its general characteristics and the method used to calculate its density and seismic wave speeds.
This process is based on an approach similar to that of \citet{Vilella2015b}.
\subsection{General description of the reference composition \label{sec21}}
Our (reference) pyrolitic composition is composed of 18\,vol\% ferropericlase (Mg,Fe)O (hereafter called Fp), 75\,vol\% Bridgmanite (Mg,Fe)(Al,Si)O$_{3}$ (hereafter called Bm), and 7\,vol\% Ca-silicate perovskite CaSiO$_{3}$ (hereafter called CaPv).
The assemblage contains 8~wt\% FeO and 3.6~wt\% Al$_{2}$O$_{3}$ to stick to a primitive mantle composition \citep[e.g.][]{McDonough1995}.
We further assume that iron in bridgmanite is 50\% ferric and 50\% ferrous, i.e., an oxidation state Fe$^{3+}$/$\sum$Fe equal to 0.5.
The Fe-Mg exchange coefficient between Bm and Fp,
\begin{equation}
K^\mathrm{Bm-Fp}= \dfrac{\biggl(\dfrac{\mathrm{Fe}}{\mathrm{Mg}}\biggr)_\mathrm{Bm}}{\biggl(\dfrac{\mathrm{Fe}}{\mathrm{Mg}}\biggr)_\mathrm{Fp}},
\label{Kd}
\end{equation}
is assumed to be constant and equal to 0.5 \citep{Piet2016}.
Using this value, combined with the assumed FeO content, we can obtain the proportion $x_{fp}$ of FeO in ferropericlase, ($1-x_{fp}$)MgO-($x_{fp}$)FeO, and the proportion $x_{Bm}$ of iron in Bm (Supplementary Material).
To determine the proportion of the different components in Bm, we assume that ferrous iron enters into Bm as FeSiO$_{3}$, while ferric iron enters into Bm as FeAlO$_{3}$. 
If there is an excess of Fe$^{3+}$, Fe$_{2}$O$_{3}$ enters into Bm, whereas if there is an excess of Al, Al$_{2}$O$_{3}$ enters into Bm.
Finally, we include the effect of the Fe$^{2+}$ spin state transition in ferropericlase following \citet{Vilella2015b}.
A more detailed description of the model is available in Supplementary Material.

\subsection{Density of the rock assemblage}
We first estimate the isothermal bulk modulus at ambient conditions $K_{T0}$ of pure end-members assuming a Voigt-Reuss-Hill average and using measurements of polycrystal (Supplementary Table 1).
Using these derived values combined to the proportion of the different components (described in section~\ref{sec21} and Supplementary Material), we calculate the $K_{T0}$ of the three minerals assuming again a Voigt-Reuss-Hill average.
We then use the Mie-Gr\"uneisen-Debye equation of state \citep{Jackson1996} to calculate the ratio V$_{0}$/V (or equivalently $\rho / \rho_{0}$) of each mineral,
\begin{equation}
\begin{split}
P= \dfrac{3K_{T0}}{2} \biggl[\biggl(\dfrac{V_{0}}{V}\biggr)^{7/3}-\biggl(\dfrac{V_{0}}{V}\biggr)^{5/3}\biggr]\hspace*{2cm} \\
\biggl\{ 1-\frac{3}{4}(4-K_{T0}')\biggl[\biggl(\dfrac{V_{0}}{V}\biggr)^{2/3}-1\biggr]\biggr \} + \Delta P_{th},
\end{split}
\label{EOS}
\end{equation}
where $\Delta P_{th}$ is the thermal pressure and the subscript zero indicates ambient conditions for volume $V_{0}$, temperature $T_{0}$, isothermal bulk modulus $K_{T0}$ and its pressure derivative $K_{T0}'$.
Values used for mineral properties are reported in Supplementary Table 1 and 2. 
To obtain the density ($\rho$) of each mineral we now only need to calculate their density at ambient conditions ($\rho_{0}$).
For this, we estimate V$_{0}$ for each mineral following a procedure similar to that for $K_{T0}$, except that we use a simple arithmetic mean instead of a Voigt-Reuss-Hill average.
Finally, the density of the rock assemblage is obtained by calculating the arithmetic mean of the density for the three minerals considered.
\subsection{Seismic wave speed of the rock assemblage}
The calculation of the bulk sound speed,
\begin{equation}
V_{\phi} = \sqrt{\frac{K_{s}}{\rho}},
\end{equation}
requires only the determination of the isentropic bulk modulus $K_{s}$ that can be obtained from the isothermal bulk modulus $K_{T}$,
\begin{equation}
K_{s} = K_{T}(1+\alpha \gamma T),
\label{Ks}
\end{equation}
where $\alpha$ is the thermal expansion coefficient and $\gamma$ the Gr\"uneisen parameter.
All the parameters in the eq.~\ref{Ks} can be calculated using the formalism of the Mie-Gr\"uneisen-Debye equation of state(see Supplementary Material for more details), so that the calculation of $V_{\phi}$ is straightforward.

The determination of $V_{S}$ and $V_{P}$ is however more complicated because they further depend on the shear modulus ($G$),
\begin{equation}
V_{S} = \sqrt{\frac{G}{\rho}}, \hspace{1.5cm} V_{P} = \sqrt{\frac{K_{s}+(4/3)G}{\rho}}.
\label{VsVp}
\end{equation}
We calculate $G$ in two steps.
First, we use an available equation of state to account for the pressure dependence \citep[e.g.,][]{Bina1992},
\begin{equation}
G(T = 300~K, P) = \biggl(\dfrac{V_{0}}{V}\biggr)^{5/3} \biggl[G_{0} + 0.5\biggl[1-\biggl(\dfrac{V_{0}}{V}\biggr)^{2/3} \biggr]\, \biggl(5G_{0}-3G'_{0}K_{T0} \biggr)  \biggr],
\label{G1}
\end{equation}
with $G_{0}$ and $G'_{0}$ being the shear modulus and its pressure derivative at ambient conditions.
Second, the temperature dependence is calculated following,
\begin{equation}
G(T, P) = G(T = 300~K, P) + dG/dT (T - 300),
\label{G2}
\end{equation}
where $dG/dT$ is the temperature derivative of the shear modulus.
Values of $dG/dT$ are given by laboratory experiments, while values of $G_{0}$ and $G'_{0}$ are estimated using measurements of polycrystal following a similar procedure that for the isothermal bulk modulus (Table~\ref{ShearModulusTab}).
 
It is important to note that the determination of the shear modulus is subject to much larger uncertainties than the determination of the density and bulk modulus.
First, the experimental measurements are more challenging.
Less data are available, and measurements are less robust.
As such, we also consider results from ab-initio calculations for the shear modulus of Al-bearing bridgmanite \citep{Shukla2015, Shukla2016}.
Second, eqs.~\ref{G1} and \ref{G2} used to extrapolate measurements are also less accurate.
In particular, we use the temperature derivative of the shear modulus to estimate its temperature dependence (eq.~\ref{G2}), while it is likely to induce large uncertainties, since this temperature derivative is probably not constant with pressure \citep{Shukla2016}.

The procedure developed here allows to calculate the density, S-wave velocity, P-wave velocity and bulk sound speed as a function of pressure and temperature.
The results for our reference composition are displayed in Supplementary Material.
Interestingly, the same procedure can be applied to a large diversity of compositions, especially to compositions that may be relevant for LLSVPs.
In the following section, we describe all the potential compositions investigated, with a particular emphasis on those explaining the seismic signature of LLSVPs.

\section{Potential compositions of LLSVPs \label{secLLSVP}}
\subsection{Range of compositions investigated}
The purpose of this work is not to test specific LLSVPs compositions or natures, but to follow an approach with no preconception on their compositions.
However, it is in practice impossible to consider all the minerals that potentially exist at the bottom of the mantle.
The compromise proposed here is to focus only on rock assemblages consisting of the three minerals expressed in a pyrolitic composition (Bm, CaPv, Fp), while varying all compositional parameters within this assemblage. 
By varying the proportions of SiO$_{2}$, MgO, CaO, Al$_{2}$O$_{3}$, and FeO, we can vary the phase proportions of Bm, CaPv, and Fp, as well as the iron and aluminium content in these phases along with the oxidation state.
For the sake of simplicity, we chose to describe these compositional parameters as the volumetric phase proportions of Bm and CaPv (Fp is bound to the others as the sum Fp+Bm+CaPv has to be 100\%), Al$_{2}$O$_{3}$ content which is incorporated in Bm, FeO content which partitions between Bm and Fp according to the partition coefficient (Eq.~\ref{Kd}), and the oxidation state (Fe$^{3+}$/$\sum$Fe in Bm).
Each of these parameters is varied in a systematic way within the lower and upper bounds listed in Table~\ref{TabParam}.

Finally, we prescribed the pressure and temperature conditions relevant for both the far-field mantle and LLSVPs.
We consider a far-field mantle with a pressure of $P = 130$~GPa, corresponding to the lowermost mantle, and a temperature of $T = 3000$~K.
For the sake of simplicity, the same pressure $P$ is considered for LLSVPs, while, because LLSVPs temperature is unknown (and likely to vary laterally), we allow their temperature $T + \Delta\, T$ to vary.
The temperature contrast ($\Delta\, T$) between far-field mantle and LLSVPs stands as the sixth parameters of our model (Table~\ref{TabParam}).
In the following, this six-dimensional space is sampled uniformly giving an ensemble composed by more than 81 millions models to investigate.

\subsection{Influence of explored parameters on seismic properties}
\begin{figure}
\includegraphics[width=0.99\textwidth]{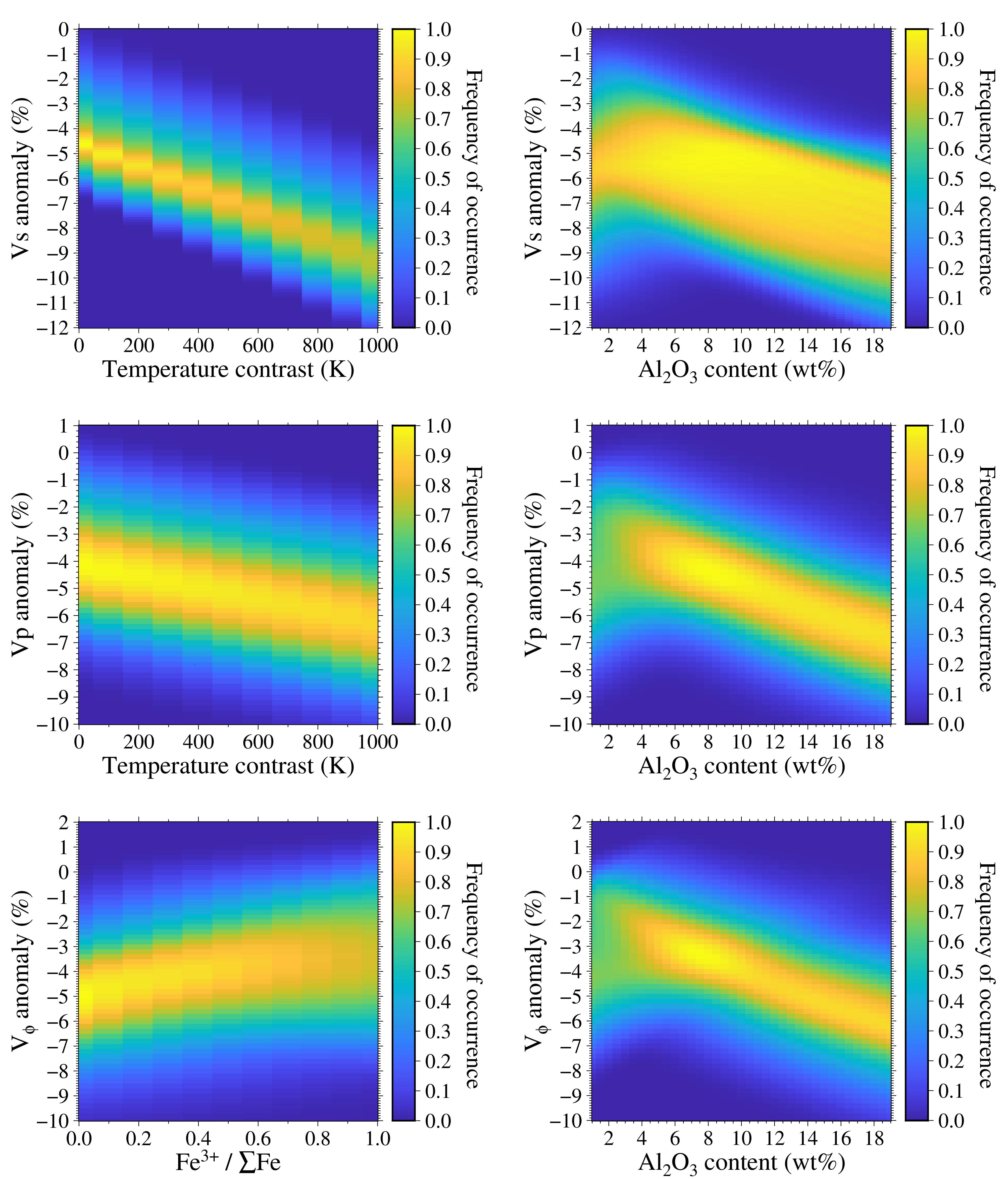}
\caption{\label{ResultsOutput} 2D histograms showing the distribution of all the models as a function of one input parameter (x-axis) and one output (y-axis). The effect of only two parameters, corresponding to the dominant ones, are reported for each seismic wave speed. Other histograms are shown in Supplementary Material.}
\end{figure}

Before identifying and discussing potential compositions of LLSVPs, it is important to characterize the effects of the six parameters on density and seismic wave speeds.
Because of the relatively large number of models and input parameters, we mainly present results using 2D histograms showing the distribution of all models as a function of one input parameter and one output ($V_{P}$, $V_{S}$, $V_{\phi}$, $\rho$).
The main interest of these histograms is to reveal the trends between input parameters and outputs.
Narrow distributions and large variations indicate that the input parameter is likely the dominant parameter affecting the output.
By contrast, wide distributions without clear variations suggest that the input parameter does not substantially affect the output.
We plotted the 2D histograms for each output as a function of all 6 parameters.
Figure~\ref{ResultsOutput} shows the histograms corresponding to the dominant parameters (other histograms are shown in Supplementary Material).

The $V_{S}$ anomaly is dominantly affected by the temperature contrast, with decreasing $V_{S}$ as the temperature contrast increases.
Note that the $V_{S}$ anomaly decreases by about 1\% for only a 200~K increase, which stands as a very strong effect.
The Al$_{2}$O$_{3}$ content also plays a role, with a $\sim$2\% decrease of $V_{S}$ for a 10wt\% increase of the Al$_{2}$O$_{3}$ content.
However, large change in Al$_{2}$O$_{3}$ content is required to impact $V_{S}$, so that temperature variations appear to be a more straightforward mechanism to produce $V_{S}$ anomalies in the lower mantle.
The $V_{P}$ anomaly exhibits a similar behavior as the $V_{S}$ anomaly, except that the Al$_{2}$O$_{3}$ content has a slightly larger impact than temperature.
The $V_{\phi}$ anomaly is mainly affected by the Al$_{2}$O$_{3}$ content, with $V_{\phi}$ decreasing as Al$_{2}$O$_{3}$ content increases, and to a lesser extent by the oxidation state with $V_{\phi}$ increasing with Fe$^{3+}$/$\sum$Fe.
Finally, density is, as expected, mainly affected by the FeO content, with increasing $\rho$ as FeO content increases.

From a more general perspective, one may note that the three seismic wave speeds exhibit similar behaviors with variations of input parameters (Supplementary Figure 2--4).
In particular, seismic wave speeds are decreasing with increasing FeO content.
The effects of the temperature contrast and FeO content are therefore in agreement with the findings of previous studies \citep{Samuel2005, Deschamps2012b}.
Perhaps more surprising is the nonlinear effect of the Al$_{2}$O$_{3}$ content (figure~\ref{ResultsOutput}).
For Al$_{2}$O$_{3}$ content up to 4~wt\%, seismic anomalies slightly increases and decreases for larger contents.
This nonlinear effect shows that it may not be fully appropriate to consider a constant seismic sensitivity throughout the explored parameter range as it has been done for simpler mineralogical compositions \citep{Deschamps2012b}.
Finally, we note that the FeO contents investigated here (up to 28wt\%) are much higher than in previous studies (up to 14wt\%).
Our results suggest that adding Al$_{2}$O$_{3}$ in the composition changes importantly the properties of the rock assemblage and allows to incorporate a much larger proportion of iron while maintaining a reasonable density.
The incorporation of alumina is therefore crucial to investigate realistic mantle compositions and to obtain more robust results. 
\subsection{Constraints on LLSVPs}

It is difficult to determine unambiguously the precise characteristics of LLSVPs.
First, these properties are likely to vary laterally, because of lateral variations of temperature and composition \citep{Ballmer2016}.
Second, different tomographic models suggest different properties depending on the dataset and methods used.
Third, tomographic models are generally obtained through a linearized and regularized inversion that includes damping and smoothing, so that the actual seismic signature of LLSVPs may be underestimated.

As a consequence, we choose here to select compositions satisfying rather conservative conditions:
\begin{itemize}
\item $V_{S}$ anomaly compared to the far-field mantle between -7\% and -2\%.
\item $V_{P}$ anomaly compared to the far-field mantle lower than -0.5\%.
\item Positive $V_{\phi}$ anomaly compared to the far-field mantle.
\end{itemize}
Note that constraining the $V_{\phi}$ anomaly is challenging because it relies on the determination of $V_{S}$ and $V_{P}$ that are usually imaged with separate tomographic inversions at different resolution and with different levels of uncertainties. 
Nevertheless, seismic tomography \citep{Masters2000, Houser2008} and normal mode data \citep{Ishii1999, Trampert2004} suggest a positive anomaly.
We thus follow a conservative approach by applying this condition without setting an amplitude for the anomaly.
In addition to these seismic constraints, we also use geodynamical constraints and restrict our study to models with a compositional density difference between 2\% and 2.8\% compared to the far-field mantle.
This condition is suggested by the density difference required to produce chemically distinct reservoirs with the shape of LLSVPs in geodynamic models \citep[e.g.,][]{Li2014}.
For lower compositional density differences, the material is entrained and mixed by mantle convection so that reservoirs are not stable.
For larger compositional density differences, gravity segregation occurs and reservoirs tend to form layers without large topography.
Among the $\sim$81 millions models studied, only about 79000 satisfy these conditions.
Figure~\ref{ResultsOutput} suggests that most of cases are compatible with the $V_{P}$ and $V_{S}$ anomalies deduced from seismic observations.
Hence, constraining potential compositions of LLSVPs requires constraints on $V_{\phi}$ and $\rho$.
Interestingly, as discussed in section~\ref{secLLSVP}, our calculation of $V_{\phi}$ and $\rho$ is much more robust than of $V_{P}$ and $V_{S}$.
Therefore, uncertainties on the determination of the shear modulus, and thus on $V_{P}$ and $V_{S}$, do not affect significantly our results.
Note that from a seismological point of view, it is the opposite, since $V_{P}$ and $V_{S}$ are much better constrained than bulk velocity and density.
In the remainder of this study, we focus on the 79000 successful models.
Note that the properties of these models are available for academic purposes \citep{Vilella2020}.

\subsection{Characteristics of LLSVPs \label{secPossible}}
\begin{figure}
\includegraphics[width=0.99\textwidth]{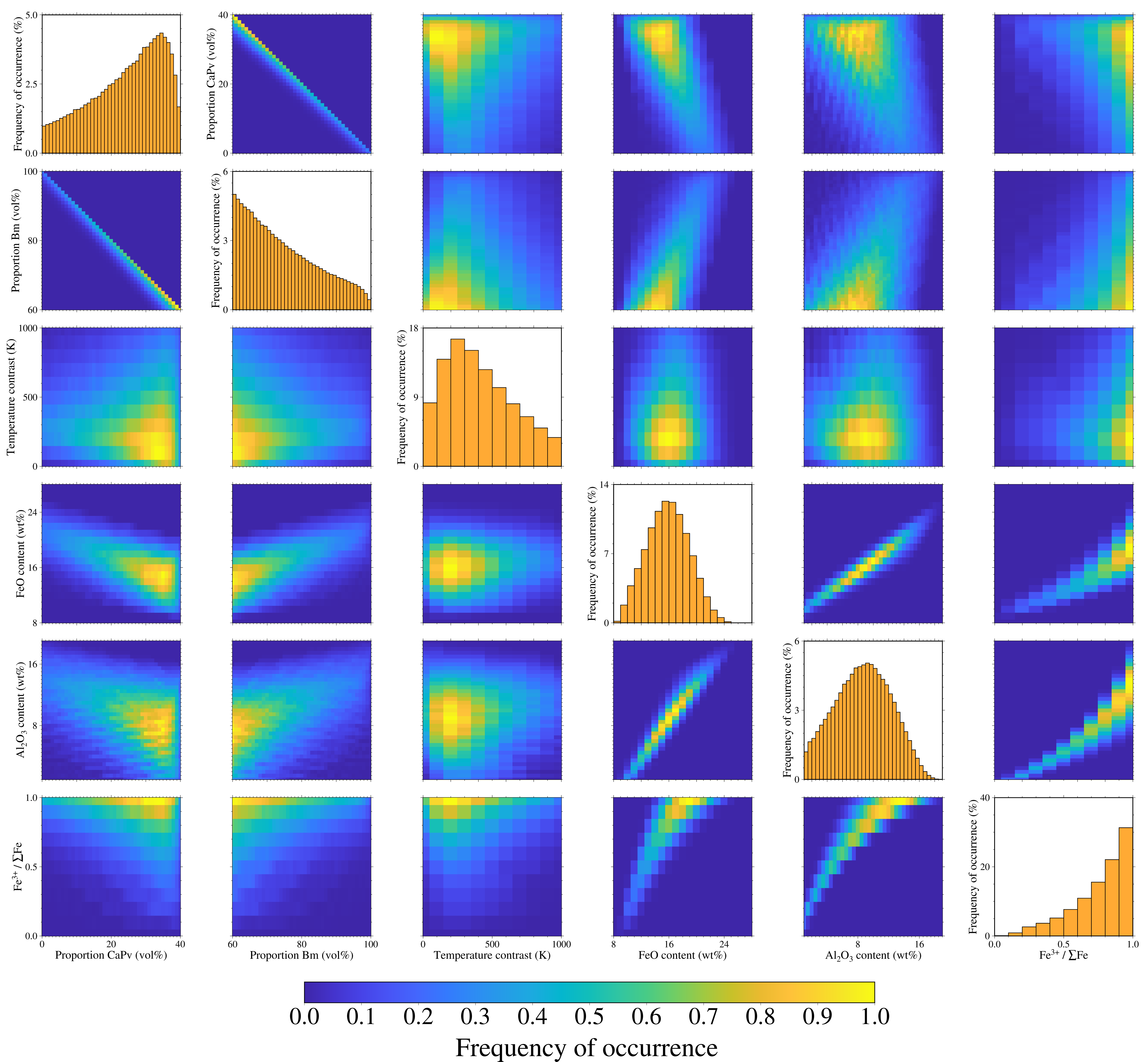}
\caption{\label{Hists} Histograms and 2D-histograms showing the distribution of the successful models for all the input parameters (see text for more details).}
\end{figure}

The 1D and 2D-histograms showing the distribution of the successful models are reported in figure~\ref{Hists} for the six parameters.
It is important to note that these distributions are not indicative of the probability of a composition to explain LLSVPs, as changing the sampling of the input parameters (i.e., lower and upper bounds) may change substantially the observed distributions.
Moreover, the sampling of the parameter space is intrinsically uneven because of the overall decreasing number of possible composition with increasing proportion of Bm.
A simple way to explain this is that in the extreme case of 100\% of Bm, CaPv is of course absent, whereas for a proportion of Bm equal to 60\%, the proportion of CaPv+Fp is equal to 40\%, so that the proportion of CaPv can take any values between 0 and 40\%.
This uneven sampling tends to increase the frequency of occurrence for low proportion of Bm or CaPv.
It is therefore important to focus only on the trends between the different parameters and on the conditions for which no successful composition have been found, which are both robust features.

A thorough analysis of the results is however difficult because of the large number of parameters and successful models.
To overcome this issue, we can try to reduce the number of parameters by identifying existing trade-offs.
For instance, there is a strong anti-correlation between the proportion of Bm and CaPv (figure~\ref{Hists}), i.e., an increase in the proportion of Bm have to be balanced by a decrease in CaPv (and vice versa).
More specifically, figure~\ref{Hists} shows that the proportion of Fp should remain almost constant and is typically $<$6\%, i.e., LLSVPs should be characterized by a much lower proportion of Fp than the far-field mantle.
Because Fp has a lower $V_{\phi}$ than other minerals, higher proportions of Fp are characterized by a negative $V_{\phi}$ anomaly, in contradiction with available constraints, and are therefore unlikely.
Furthermore, this anti-correlation also suggests the existence of a linear relationship between the proportion of Bm and CaPv, i.e., these two parameters are not independent and one of them can be chosen as a representative for the two.
In order to identify and quantify all the existing trade-offs, we apply a principal component analysis (PCA) to our dataset.

\subsubsection{Principal Component Analysis}
\begin{figure}
\includegraphics[width=0.99\textwidth]{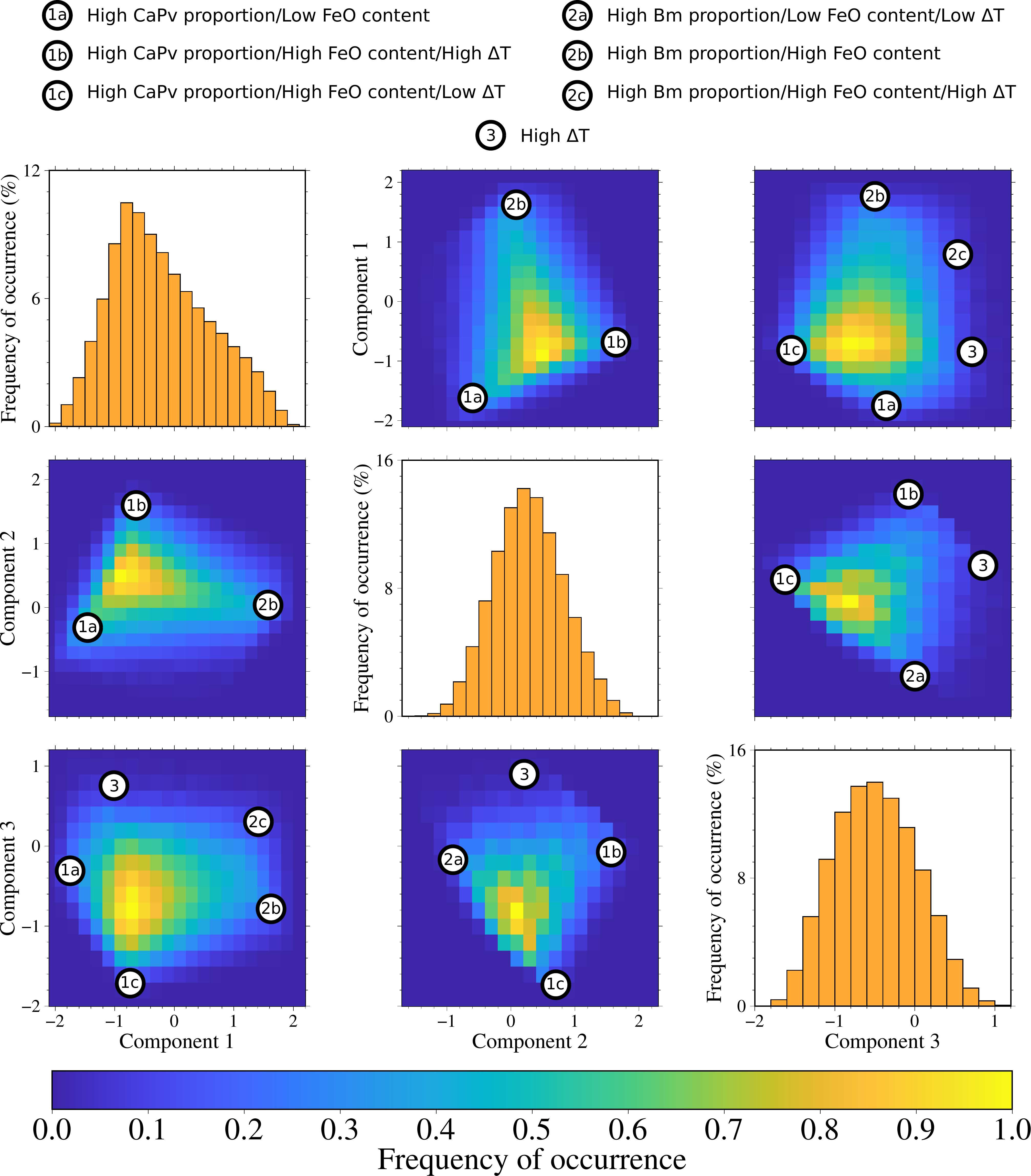}
\caption{\label{HistEigen} Histograms and 2D-histograms showing the distribution of the successful models projected in the principal components space (eigenvectors 1--3 in Table~\ref{TabPCA}). As the principal components space has been centred and normalized, the interpretation of the results requires a rescaling. We therefore provide (above the figure) the characterization of the different end-members cases shown by the 2D-histograms. Note that high CaPv proportions are associated with low Bm proportions (and vice versa), while high (low) FeO contents are associated with high (low) Al$_{2}$O$_{3}$ contents and oxidation states.}
\end{figure}

The PCA provides a projection of a dataset in a new orthogonal coordinate system where the vectors, called components, are linear combinations of the input parameters.
In this approach, the new vectors are successively built such that they best explain the variability of the dataset.
The relative importance of each component is given by their corresponding eigenvalue.
If one or several eigenvalues are much lower than others, then all the variability can be explained by a lower number of parameters, indicating the existence of one or several trade-offs between the input parameters.
The components with a high eigenvalue can be analyzed to highlight the underlying trade-offs, and in turn reduces the number of independent parameters.
The six components of our dataset with their eigenvalues are reported in Table~\ref{TabPCA}.
Because the eigenvalues 1-3 are much higher than the eigenvalues 4-6, the components 4-6 can be viewed as negligible and the components 1-3 are the principal components.
The dimension of our parameter space can therefore be reduced from 6 to 3 independent parameters.
To verify this, we plotted our dataset in the orthogonal system formed by the eigenvectors 1-3 (figure~\ref{HistEigen}).
In this coordinate system, the distributions do not exhibit any trade-offs confirming that all the information have been extracted by the PCA.
We now conduct a detailed analysis of the principal components in order to identify these trade-offs.

The first component is dominated by the opposite contribution from the proportion of Bm (0.600) and CaPv (-0.591).
Note that, although weaker, a similar behavior can be observed in the components 2 and 3.
This corresponds to the anti-correlation between CaPv and Bm discussed in section~\ref{secPossible}.
The second component is dominated by the contribution of the Al$_{2}$O$_{3}$ content and oxidation state, while the temperature contribution is also significant.
By comparing the three principal components, one may see that the FeO content, Al$_{2}$O$_{3}$ content and oxidation state are always correlated.
These correlations can also be seen from the 2D-histograms displayed in figure~\ref{Hists} and from the end-members identified in figure~\ref{HistEigen}.
Interestingly, these three elements are required to form the FeAlO$_{3}$ component in Bm.
We therefore postulate that the FeAlO$_{3}$ component in Bm has a better ability to explain observations than other iron-bearing components, which causes a robust correlation between the FeO content, Al$_{2}$O$_{3}$ content and oxidation state.
To support this point, we have verified the low proportion of Al$_{2}$O$_{3}$ and Fe$_{2}$O$_{3}$ components in all the successful cases (see Supplementary Figure 5).
Furthermore, increasing the proportion of the FeAlO$_{3}$ component has the direct effect of increasing density.
This component may thus be interpreted as a density shift vector, while the third component affects mainly the temperature.

From the analysis of the end-members in figure~\ref{HistEigen} and the principal components, we postulate that all the successful cases can be separated in two end-member compositions: one CaPv-rich and one Bm-rich.
For both composition, we applied a PCA and found the same trade-offs as for the full set of successful models, e.g., the correlation between FeO content, Al$_{2}$O$_{3}$ and  Fe$^{3+}$/$\sum$Fe.
We now study in more detail these two end-member compositions beginning with the Bm-rich case.

\subsubsection{End-member composition: Bm-rich}

Building on the results of the PCA, we apply an additional condition to our successful models, that is the Bm proportion should be larger than 90vol\%.
The purpose of this condition is to extract the Bm-rich end-member composition from the tally of successful models.
We have selected this specific proportion because it is high enough to be restrictive, while providing a number of cases ($\sim$9000) sufficient for the data analysis.  
In the following, the results are displayed using histograms of the successful models as a function of two input parameters.
As suggested by the PCA, we focus on the effect of $\Delta\, T$ and the FeO content, which can be viewed as a representative parameter for the  Al$_{2}$O$_{3}$ content and oxidation state.

\begin{figure}
\includegraphics[height=0.99\textwidth]{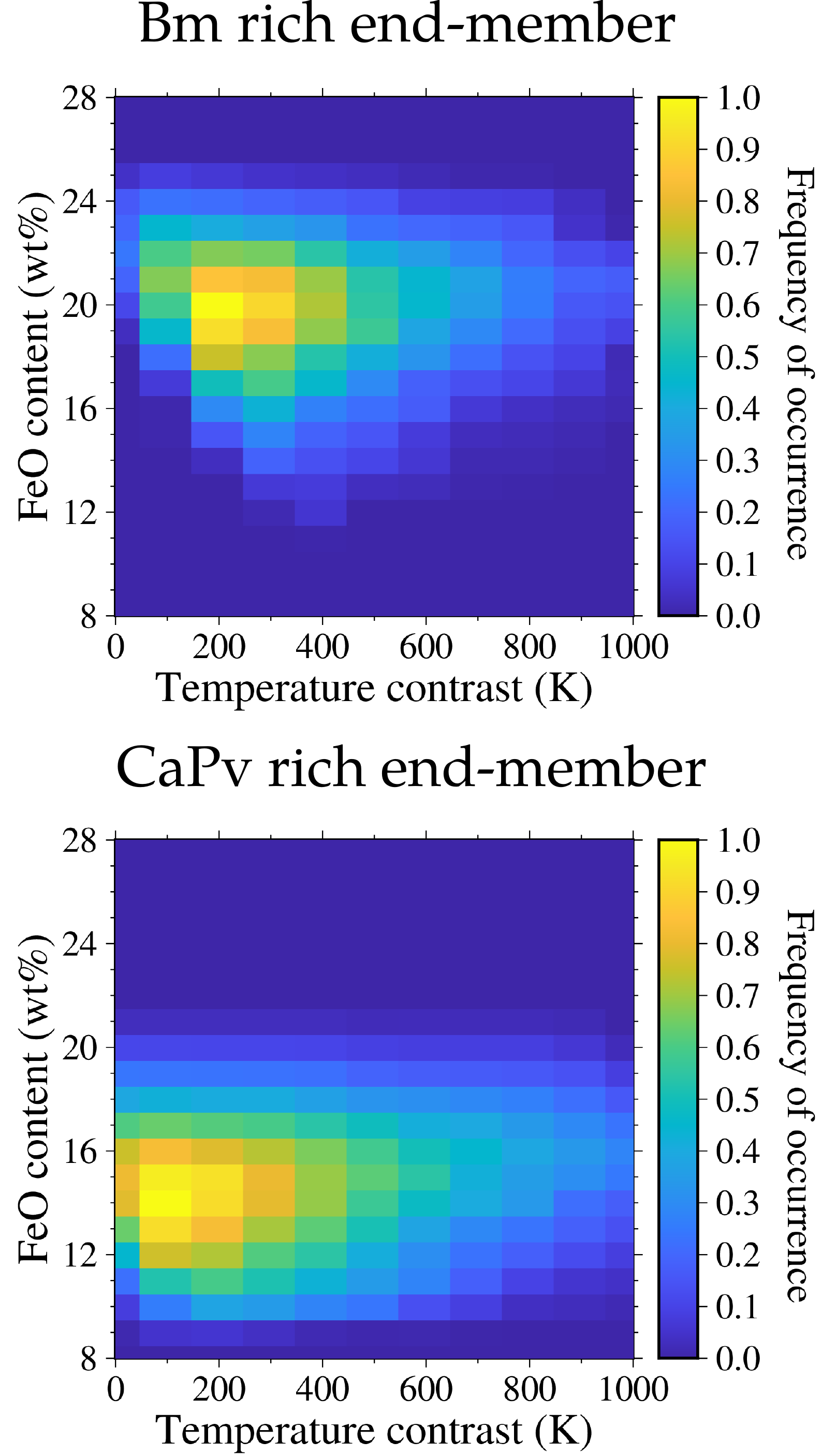}
\caption{\label{HistEM} 2D histograms showing the distribution of the models corresponding to the Bm-rich case (top panels) and CaPv-rich case (bottom panels) as a function of the temperature contrast and FeO content.}
\end{figure}

Figure~\ref{HistEM}a shows the distribution of the Bm-rich models as a function of $\Delta\, T$ and FeO content.
For these models, a large range of temperature is possible, while larger temperatures seem slightly more likely for FeO content $\sim$21wt\%.
Although not fully robust, observations may be easier to explain with temperature contrast lower than 500~K.
FeO contents are typically $>$14wt\% with values up to 25wt\%.
As predicted, Al$_{2}$O$_{3}$ content (typically $>$8wt\%, not shown here) and oxidation state (typically larger than 0.6, not shown here) are also high.
Compositions more enriched in Bm seem to slightly favor larger FeO contents, Al$_{2}$O$_{3}$ contents and oxidation states while involving slightly lower temperature contrasts.

\subsubsection{End-member composition: CaPv-rich}

We follow a similar procedure for the CaPv-rich end-member composition by selecting only the successful cases with a proportion of CaPv larger than 30vol\%.
A total of 32000 cases was obtained from the 79000 initial successful cases.
The main results are displayed in figure~\ref{HistEM}b.
As for the Bm-rich case, a wide range of temperature is possible, but observations seem easier to explain with a temperature contrast lower than 500~K.
Note that the distribution is now slightly more scattered than in the Bm-rich case and does not exhibit any specific trend with increasing proportion of CaPv, this result being valid for all the parameters.
The FeO content is typically between 10wt\% and 21wt\%, the Al$_{2}$O$_{3}$ content between 3wt\% and 13wt\%, and the oxidation state larger than 0.3.

\section{Discussion}

\subsection{Summary of the results}

We investigated a large range of models for LLSVPs with varying FeO content, Al$_{2}$O$_{3}$ content, proportion of CaPv, proportion of bridgmanite, oxidation state Fe$^{3+}$/$\sum$Fe, and temperature contrast with respect to the far-field mantle $\Delta\, T$.
We found that constraints on density and bulk sound speed of LLSVPs are by far the most useful observations to identify the potential composition of these structures.
By contrast, the low $V_{S}$ and $V_{P}$ observed are easily achieved as long as LLSVPs are a few hundreds Kelvin hotter than the far-field mantle.
Two robust results are that the proportion of Fp should be low ($<$6vol\%), and that the formation of Fe$^{3+}$-bearing bridgmanite is much more favorable than other iron-bearing components.
As a consequence, we observed a clear trade-off between FeO content, Al$_{2}$O$_{3}$ content and Fe$^{3+}$/$\sum$Fe.
To describe the large number of models, we identified two end-member compositions: one Bm-rich and one CaPv-rich.
For each end-member, a large range of temperature is possible, although it seems easier to explain seismic observations with low temperature contrasts ($<$500~K).
Furthermore, we found fairly high FeO and Al$_{2}$O$_{3}$ content, up to 25wt\% and 19wt\%, respectively.
It is important to note that a full range of compositions is possible between these two end-members, so that, from a more general perspective, a very large range of parameters is possible for LLSVPs.
Interestingly, temperature excess deduced from normal modes data \citep{Trampert2004} and attenuation \citep{Deschamps2019} are in good agreement with our findings, while the high Al$_{2}$O$_{3}$ content is consistent with the super-chondritic Ca/Al ratio measured for the accessible Earth \citep[e.g.,][]{Walter2004}.

\subsection{Limitations and potential developments}
\subsubsection{Effects of post-perovskite}
Our results rely on the selected lower mantle mineralogy and on the thermo-elastic properties of these minerals.
More specifically, other minerals that the three used in our model may be present in the lowermost mantle.
In particular, we did not account for the possible presence of post-perovskite \citep[PPv, see][for a review]{Shim2008}, a high-pressure phase of bridgmanite.
PPv is characterized by a slightly higher density, lower bulk sound speed, and higher shear-wave speed than Bm \citep[see][for a compilation]{Cobden2015}, such that its presence outside LLSVPs has been advocated as an explanation for the anti-correlation between $V_{\phi}$ and $V_{S}$-anomalies \citep[e.g.,][]{Davies2012}.
A difficulty, however, is to precisely determine the stability field of PPv, which depends on temperature and pressure \citep[e.g.,][]{Catalli2009}, but also on the amount of iron and alumina \citep[e.g.,][]{Sun2018}.
In particular, the transition pressure increases with increasing temperature and decreasing alumina and iron content.
Because of these dependencies, and following the assumed conditions and composition, PPv may or may not be present in the lowermost mantle.
In our case, the assumed far-field mantle is likely composed of a mixture of Bm and PPv.
If only the thermal effect is accounted for, the transition pressure of PPv should be too large to occur within LLSVPs.
However, if LLSVPs are strongly enriched in alumina and iron, the transition pressure of PPv may be smaller than the CMB pressure \citep{Sun2018}, i.e., PPV may be stable within LLSVPs.
In that case, PPv may be present both outside and within LLSVPs, such that its seismic signature would be reduced.
Unfortunately, available constraints concerning the effects of alumina and iron content on the properties of PPv are too scarce to allow a careful estimate of its effects.
\subsubsection{Uncertainties}
Although our reference composition is compatible with available constraints (see Supplementary Material), the precise composition of the far-field mantle remains uncertain.
The uncertainties that may affect our reference composition should therefore be properly addressed.
In that aim, we have investigated the impact of compositional variations in our reference composition.
The results show that changing the reference composition shift the range of possible compositions for LLSVPs.
For instance, increasing (decreasing) the proportion of Fp in the reference composition induces an increase (decrease) of the maximum proportion of Fp in the range of possible compositions for LLSVPs.
Nevertheless, we found that the modification of the reference composition does not impact the main conclusions of our work.
The proportion of Fp in LLSVPs has to be lower than in the reference composition, while the formation of Fe$^{3+}$-bearing bridgmanite is favored over other iron-bearing components.
Furthermore, there are uncertainties on the exact value of mineral properties, especially concerning the shear modulus, as contrasted results have been found for the elasticity of Fp \citep{Murakami2012, Wu2013}.
These uncertainties may impact our results.
As a matter of fact, using the ab-initio results of \citet{Wu2013}, instead of the experimental results of \citet{Murakami2012}, leads to LLSVPs with a maximum proportion of Fp that is slightly higher (from $\sim$6\% to $\sim$8\%).

Overall, this supports our methodology relying on the identification of the general trends in the potential composition of LLSVPs, rather than on the search of the most probable composition.
Since both the unknown parameters and observations are defined as relative to a reference lower mantle, an error in the reference mantle will only produce a bias (e.g. a systematic error) in our results.
Therefore, all the relative information between inferred parameters (trade-offs, and relative compositions) are robust features that remain mostly unaffected by reasonable uncertainties on the reference composition.  
  
\subsubsection{Additional constraints}
One way to reduce the range of plausible thermo-chemical models for LLSVPs is to use tighter constraints.
Improvement in tomographic models, leading to a precise estimate of the $V_{\phi}$ anomaly, would reduce the range of possible compositions.
Additional constraints may also be useful.
For instance, tidal tomography \citep{Lau2017} provides a constraint on the effective density difference ($\Delta \rho_{tot}$), i.e., including the thermal and compositional effects, between LLSVPs and the far-field mantle.
Using a simple model of the lower mantle composed of three 340~km thick layers, \citet{Lau2017} found that LLSVPs should be $\sim$0.60\% denser.
In comparison, the minimum $\Delta \rho_{tot}$ of our successful cases is slightly larger, at $\sim 1 \%$.
This discrepancy may be related to the underestimation of $\Delta \rho_{tot}$ in tidal tomography models due to a dilution effect, or may be indicative of a lower density difference between LLSVPs and far-field mantle than the one assumed in this work.
In both cases, constraining the effective density difference stands as a very robust way to infer the nature and composition of LLSVPs.
Supplementary constraints may further include seismic attenuation.
Assuming that seismic attenuation is mainly affected by temperature, it can be used to constrain the temperature conditions of LLSVPs \citep{Deschamps2019}.
Although it may be affected by several sources of uncertainties, it is interesting to note that in the western tip of the Pacific LLSVP, modeling of attenuation predicts a temperature excess of 350 +/- 200 K \citep{Deschamps2019}, a value consistent with the temperature excess suggested by our calculations. 
A better knowledge of mineral properties may refine the modeling of attenuation, allowing sharper constraints on the potential composition of LLSVPs, and therefore the exploration of more complex and detailed mineralogical models.
Alternatively, and as illustrated in the next section, the range of potential compositions may be reduced by considering the potential process that created these LLSVPs, and whether the produced compositions are reasonable.

\subsection{Potential origin of LLSVPs}
\subsubsection{The magma ocean hypothesis}

Some authors have postulated that the final stages of the solidification of a primitive magma ocean could be the source of LLSVPs \citep[e.g.,][]{Labrosse2007, Ballmer2017}.
However, the potential composition produced by such a scenario remains largely uncertain, mainly because the solidifying process itself is uncertain \citep[see][for more details]{Solomatov2015}.
First, different modes of crystallization are possible that may or may not produce chemically distinct reservoirs.
Second, the sequence of crystallization is highly debated and depends, in particular, on the relative behavior of the adiabatic temperature gradient and the liquidus profile \citep[e.g.,][]{Boukare2017}, which, again, is uncertain at lower mantle conditions \citep{Andrault2011, Nomura2011}.
Moreover, depending on the value of the iron partitioning coefficient between solid and melt, the crystals could be denser than melt or not \citep{Nomura2011, Andrault2012, Tateno2014, Caracas2019}.
Considering all these uncertainties, the magma ocean could solidify from the bottom and propagate to the surface (bottom-up scenario), or from the middle and propagate to both the surface and the CMB (middle-out scenario).
In the latter case, a basal magma ocean (BMO) is produced that may give birth to LLSVPs \citep{Labrosse2007}.
In the bottom-up scenario, the dense reservoir is located near the surface and should overturn to produce a stable density gradient \citep{ElkinsTanton2003}, potentially producing a basal magma ocean \citep{Ballmer2017}.

Independently of the solidification scenario, the residual liquid should be enriched in iron and alumina \citep{Tateno2014, Boujibar2016}.
A first order estimate of such an enrichment can be obtained using the simplified model proposed by \citet{Boukare2015, Boukare2018}.
In this model, the Nerst partition coefficient $D$ between melt and solid is assumed to be constant so that,
\begin{equation}
C_{s} = C_{s, 0} D \biggl( \dfrac{R^{3}_{min} - R^{3}_{max}}{R^{3}_{min} - R^{3}} \biggr)^{1-D},
\end{equation}
where $C_{s}$ is the proportion of the element in the solid, $C_{s, 0}$ the initial proportion, $R_{max}$ the Earth's radius, $R_{min}$ the CMB radius, $R$ the radius considered.
Considering that LLSVPs originate from a layer 300~km thick, we obtained a FeO content of about 28wt\% for $D = 0.6$ and above 50wt\% for $D = 0.4$.
A reasonable FeO content can therefore only be obtained for $D > 0.5$, which may require a low degree of melting \citep{Boujibar2016}.
For such low degrees of melting, alumina may be partitioned equally between melt and solid \citep{Boujibar2016}.
This stands in contradiction to the alumina enrichment suggested by our results.
Furthermore, if we consider a thinner magma ocean ($\sim$1000~km), this constraint is somewhat relaxed, but still requires low degrees of melting ($<$10\%) that seems unrealistic for the slow-cooling basal magma ocean.
Another possible issue is the oxidation state of the magma ocean, which should be initially fairly low.
However, FeO-disproportionation may induce a progressive increase of the oxidation state during the crystallization.

Alternatively, in the bottom-up scenario, the dense reservoir (highly enriched in FeO) could undergo partial melting during its overturn \citep{Ballmer2017}.
The subsequent melt-rock interactions may alter the composition of the dense reservoir, decreasing its FeO content while increasing its Al$_{2}$O$_{3}$ and SiO$_{2}$ contents.
Overall, since the style of magma-ocean freezing is poorly constrained, and element partitioning at high pressures is difficult to measure, it remains challenging to estimate precisely the composition of the produced reservoirs.

Another way to evaluate the relevance of the magma ocean origin is by looking at mineral proportion.
It has been suggested that Bm is the first mineral to crystallize, followed by Fp and CaPv \citep{Tronnes2002}.
In that case, the residual liquid of magma ocean should be enriched in CaPv and, to a lesser extent, in Fp.
However, our results suggest that a low proportion of Fp ($\sim$6vol\%) is required to explain the seismic signature of LLSVPs.
Therefore, if we further assume that the produced reservoirs should not be significantly altered during the Earth's evolution, these reservoirs should be highly enriched in SiO$_{2}$ to prevent the formation of Fp.
If one assumes that LLSVPs are related to magma ocean solidification, then an additional process \citep[e.g.,][]{Ballmer2017, Hirose2017} is required to explain the depletion in Fp.
As a consequence, we propose that the low proportion of Fp may be a more robust constraint than the FeO content to confirm or not the magma ocean origin for LLSVPs.

\subsubsection{The MORB hypothesis}

Another possible origin for LLSVPs is the accumulation of MORB materials.
At lower mantle conditions, MORB is believed to be composed of $\sim$20vol\% of CaPv, $\sim$40vol\% of Bm and other phases, such as calcium ferrite‐type phase, accounting for the remaining $\sim$40vol\% \citep{Ricolleau2010}.
In addition, MORBs cover a large compositional range \citep{Gale2013}, including substantial differences in the amount of FeO, and may further be chemically altered during subduction and related metasomatism.
Due to this strong dispersion in composition, the density and seismic signature of recycled MORBs are still debated.
For instance, numerical simulations of mantle convection \citep{Nakagawa2010b} indicate that this dispersion may affect the ability of MORBs to accumulate at the CMB, as their density contrast with respect to the far-field mantle strongly depends on their FeO content.
Taking into account the compositional dispersion to estimate the seismic signature of high pressure MORBs, \citet{Deschamps2012b} noted that LLSVPs may not be entirely composed of recycled MORB, unless their temperature exceeds that of the far-field mantle by 1500 K or more.
Recent ab initio calculations \citep{Wang2020} further pointed out that at lowermost mantle conditions, MORBs are characterized by high seismic velocities contrasting with the seismic signature of LLSVPs.
However, it cannot be excluded that LLSVP incorporate a small or moderate amount of MORB mixed with primitive material, as assumed in the Basal Melange (BAM) hypothesis \citep{Tackley2012, Ballmer2016}.
In addition, MORB can undergo melting in the subduction channel inducing a fertilization of the mantle wedge and the production of moderately-enriched mafic lithologies, which may contribute to LLSVPs.
Interestingly, experimental results on CaPv \citep{Thomson2019} are consistent with LLSVPs moderately enriched in recycled MORBs.
Because the compositions investigated here do not include all the minerals present in MORBs and have lower SiO$_{2}$ contents, our results are not able to evaluate the hypothesis that LLSVP are entirely or partially composed of recycled MORBs. 

\subsubsection{The primitive crust hypothesis}

Based on geochemical evidences, \citet{Tolstikhin2005} suggested that the subduction of a primitive crust produced just after the magma ocean solidification could be at the origin of LLSVPs.
This primitive crust should be a mixture of komatiitic crust and chondritic regolith, whose proportion and individual composition are uncertain.
Nevertheless, we can attempt to verify whether such a mixture could be compatible with our results.
Based on lunar regolith composition \citep[e.g.,][]{McKay1991}, one may postulate that terrestrial regolith should be enriched in FeO, CaO and Al$_{2}$O$_{3}$ compared to a pyrolitic composition.
Regoliths with a moderate FeO-enrichment (16wt\%) are characterized with a reasonable Al$_{2}$O$_{3}$ content ($\sim$13wt\%), while FeO-poor regolith (6wt\%) are highly enriched in Al$_{2}$O$_{3}$ ($\sim$25wt\%).
On the other hand, large compositional variations are displayed by komatiites \citep[e.g.,][]{Arndt2008}.
In particular, some komatiites are characterized by a slight to moderate enrichment in FeO, CaO and Al$_{2}$O$_{3}$.
Using appropriate compositions, we verify that a mixture of regolith and komatiite is compatible with our results.
It is however necessary to consider a large proportion of regolith (typically larger than 70\%) with moderate FeO-enrichment, otherwise the mixture has a lower FeO content than the ones suggested by our results.

\section{Concluding remarks}
Using a massive sampling of the parameter space and a principal component analysis, we find that the seismic signature of LLSVPs could be explained as long as the proportion of Fp is low, and that a certain ratio between FeO content, Al$_{2}$O$_{3}$ content and Fe$^{3+}$/$\sum$Fe is maintained.
As a result, the seismic signature of LLSVPs can be explained with a large range of compositions.
We thus tried to restrict the range of potential compositions by considering different possible sources for LLSVPs, e.g., solidification of the magma ocean or subduction of a primitive crust.
Different scenarios generate reservoirs with a compatible concentration of major elements.
However, owing to the lack of constraints available, the mineral proportions of these reservoirs are uncertain.
It is therefore difficult to evaluate the relevance of these scenarios.
\clearpage
\newpage
\begin{table}
\caption{\label{ShearModulusTab} Shear modulus ($G_{0}$) along with its pressure ($G'_{0}$) and temperature ($dG/dT$) derivative at ambient conditions for several compounds.}
\begin{tabular}{c c c c}
\hline
Compounds & $G_{0}$ (GPa) &  $G'_{0}$ &  $dG/dT$ (GPa K$^{-1}$) \\
\hline
Ferropericlase                        & -                  & -                 & -0.02$^{\text{a}}$\\   
MgO                                   & 131$^{\text{b}}$   & 1.92$^{\text{b}}$ & - \\
0.92MgO-0.08FeO (HS)                  & 113$^{\text{a}}$   & 2.15$^{\text{a}}$ & - \\
0.92MgO-0.08FeO (LS)                  & 130$^{\text{a}}$   & 2.04$^{\text{a}}$ & - \\
Bridgmanite                           & -                  & -                 & -0.02$^{\text{a}}$\\  
MgSiO$_{3}$                           & 168.2$^{\text{c}}$ & 1.79$^{\text{c}}$ & - \\
0.875MgSiO$_{3}$-0.125FeSiO$_{3}$     & 165.2$^{\text{c}}$ & 1.80$^{\text{c}}$ & - \\
0.875MgSiO$_{3}$-0.125Fe$_{2}$O$_{3}$ & 143.1$^{\text{d}}$ & 1.88$^{\text{d}}$ & - \\
0.875MgSiO$_{3}$-0.125FeAlO$_{3}$     & 156.3$^{\text{d}}$ & 1.83$^{\text{d}}$ & - \\
0.875MgSiO$_{3}$-0.125Al$_{2}$O$_{3}$ & 162.7$^{\text{d}}$ & 1.85$^{\text{d}}$ & - \\
CaSiO$_{3}$                           & 135$^{\text{e}}$   & 1.57$^{\text{e}}$ & -0.002$^{\text{e}}$\\
\hline
\end{tabular}\\
$^{\text{a}}$ \citet{Murakami2012}.\\
$^{\text{b}}$ \citet{Murakami2009}.\\
$^{\text{c}}$ \citet{Shukla2015}.\\
$^{\text{d}}$ \citet{Shukla2016}.\\
$^{\text{e}}$ Values obtained by fitting the experimental data in Table~8 of \citet{Li2006c}.\\
\end{table}
\clearpage
\newpage
\begin{table}
\caption{\label{TabParam} Range of values investigated for each compositional parameter considered in our LLSVPs model. The step value indicates the step between 2 adjacent values.}
\begin{tabular}{c c c c}
\hline
Parameter & min value &  step value & max value  \\
\hline
FeO content (wt\%)             & 8  & 1   & 28  \\
Al$_{2}$O$_{3}$ content (wt\%) & 1  & 0.5 & 19  \\
Proportion of CaPv (vol\%)     & 0  & 1   & 40  \\
Proportion of Bm (vol\%)       & 60 & 1   & 100 \\
Fe$^{3+}$/$\sum$Fe             & 0  & 0.1 & 1   \\
$\Delta\, T$ (K)               & 0  & 100 & 1000\\
\hline
\end{tabular}\\
\end{table}
\clearpage
\newpage
\begin{table}
\caption{\label{TabPCA} Results of the principal component analysis (PCA) applied to our set of successful models. Note that the data have been centered and normalized before analysis, so that the results are here shown in a dimensionless form.}
\begin{tabular}{l c c c c c c}
            & \multicolumn{6}{c}{Components} \\
            & 1 & 2 & 3 & 4 & 5 & 6 \\
\hline
Eigenvalues             & 0.681  & 0.296  & 0.262  & 0.023  & 0.003  & 0.001 \\
\hline
Proportion CaPv         & -0.591 & 0.314  & -0.187 & -0.267 & -0.147 & 0.651\\
Proportion Bm           & 0.600  & -0.307 & 0.200  & -0.083 & -0.255 & 0.658\\
$\Delta\, T$            & 0.040  & 0.581  & 0.813  & -0.004 & 0.008  & -0.010\\
FeO content             & 0.306  & 0.217  & -0.178 & -0.406 & 0.803  & 0.137\\
Al$_{2}$O$_{3}$ content & 0.339  & 0.395  & -0.301 & -0.528 & -0.518 & -0.302\\
Fe$^{3+}$/$\sum$Fe      & 0.283  & 0.516  & -0.377 & 0.692  & -0.012 & 0.180\\
\hline
\end{tabular}\\
\end{table}
\clearpage
\newpage
\clearpage
\newpage
\clearpage
\newpage
\clearpage
\newpage
\clearpage
\newpage
\section*{Acknowledgments}
This research was funded by JSPS KAKENHI Grant Number JP19F19023.
TB is funded by the European Union's Horizon 2020 research and innovation program under Grant Agreement No. 716542.
\section*{Reference}
\bibliography{/home/kenny/Documents/Work/Article/biblio}

\newcommand{\noop}[1]{}
\begin{thebibliography}{61}
\expandafter\ifx\csname natexlab\endcsname\relax\def\natexlab#1{#1}\fi
\providecommand{\url}[1]{\texttt{#1}}
\providecommand{\href}[2]{#2}
\providecommand{\path}[1]{#1}
\providecommand{\DOIprefix}{doi:}
\providecommand{\ArXivprefix}{arXiv:}
\providecommand{\URLprefix}{URL: }
\providecommand{\Pubmedprefix}{pmid:}
\providecommand{\doi}[1]{\href{http://dx.doi.org/#1}{\path{#1}}}
\providecommand{\Pubmed}[1]{\href{pmid:#1}{\path{#1}}}
\providecommand{\bibinfo}[2]{#2}
\ifx\xfnm\relax \def\xfnm[#1]{\unskip,\space#1}\fi
%Type = Article
\bibitem[{Andrault et~al.(2011)Andrault, Bolfan-Casanova, Lo~Nigro, Bouhifd,
  Garbarino and Mezouar}]{Andrault2011}
\bibinfo{author}{Andrault, D.}, \bibinfo{author}{Bolfan-Casanova, N.},
  \bibinfo{author}{Lo~Nigro, G.}, \bibinfo{author}{Bouhifd, M.A.},
  \bibinfo{author}{Garbarino, G.}, \bibinfo{author}{Mezouar, M.},
  \bibinfo{year}{2011}.
\newblock \bibinfo{title}{{Solidus and liquidus profiles of chondritic mantle:
  Implication for melting of the Earth across its history}}.
\newblock \bibinfo{journal}{Earth and Planetary Science Letters}
  \bibinfo{volume}{304}, \bibinfo{pages}{251--259}.
\newblock \DOIprefix\doi{10.1016/j.epsl.2011.02.006}.
%Type = Article
\bibitem[{Andrault et~al.(2012)Andrault, Petitgirard, Lo~Nigro, Devidal,
  Veronesi, Garbarino and Mezouar}]{Andrault2012}
\bibinfo{author}{Andrault, D.}, \bibinfo{author}{Petitgirard, S.},
  \bibinfo{author}{Lo~Nigro, G.}, \bibinfo{author}{Devidal, J.L.},
  \bibinfo{author}{Veronesi, G.}, \bibinfo{author}{Garbarino, G.},
  \bibinfo{author}{Mezouar, M.}, \bibinfo{year}{2012}.
\newblock \bibinfo{title}{{Solid--liquid iron partitioning in Earth's deep
  mantle}}.
\newblock \bibinfo{journal}{Nature} \bibinfo{volume}{487},
  \bibinfo{pages}{354--357}.
\newblock \DOIprefix\doi{10.1038/nature11294}.
%Type = Book
\bibitem[{Arndt et~al.(2008)Arndt, Lesher and Barnes}]{Arndt2008}
\bibinfo{author}{Arndt, N.}, \bibinfo{author}{Lesher, M.C.},
  \bibinfo{author}{Barnes, S.J.}, \bibinfo{year}{2008}.
\newblock \bibinfo{title}{{Komatiite}}.
\newblock \bibinfo{publisher}{{Cambridge University Press}}.
%Type = Article
\bibitem[{Ballmer et~al.(2017)Ballmer, Louren\c{c}o, Hirose, Caracas and
  Nomura}]{Ballmer2017}
\bibinfo{author}{Ballmer, M.D.}, \bibinfo{author}{Louren\c{c}o, D.L.},
  \bibinfo{author}{Hirose, K.}, \bibinfo{author}{Caracas, R.},
  \bibinfo{author}{Nomura, R.}, \bibinfo{year}{2017}.
\newblock \bibinfo{title}{{Reconciling magma-ocean crystallization models with
  the present-day structure of the Earth's mantle}}.
\newblock \bibinfo{journal}{Geochemistry, Geophysics, Geosystems}
  \bibinfo{volume}{18}, \bibinfo{pages}{2785--2806}.
\newblock \DOIprefix\doi{10.1002/2017GC006917}.
%Type = Article
\bibitem[{Ballmer et~al.(2016)Ballmer, Schumacher, Lekic, Thomas and
  Ito}]{Ballmer2016}
\bibinfo{author}{Ballmer, M.D.}, \bibinfo{author}{Schumacher, L.},
  \bibinfo{author}{Lekic, V.}, \bibinfo{author}{Thomas, C.},
  \bibinfo{author}{Ito, G.}, \bibinfo{year}{2016}.
\newblock \bibinfo{title}{{Compositional layering within the large low
  shear-wavevelocity provinces in the lower mantle}}.
\newblock \bibinfo{journal}{Geochemistry, Geophysics, Geosystems}
  \bibinfo{volume}{17}, \bibinfo{pages}{5056--5077}.
\newblock \DOIprefix\doi{10.1002/2016GC006605}.
%Type = Article
\bibitem[{Bina and Helffrich(1992)}]{Bina1992}
\bibinfo{author}{Bina, C.R.}, \bibinfo{author}{Helffrich, G.R.},
  \bibinfo{year}{1992}.
\newblock \bibinfo{title}{{Calculation of Elastic Properties from Thermodynamic
  Equation of State Principles}}.
\newblock \bibinfo{journal}{Annual Review of Earth and Planetary Sciences}
  \bibinfo{volume}{20}, \bibinfo{pages}{527--552}.
\newblock \DOIprefix\doi{10.1146/annurev.ea.20.050192.002523}.
%Type = Article
\bibitem[{Boujibar et~al.(2016)Boujibar, Bolfan-Casanova, Andrault, Bouhifd and
  Trcera}]{Boujibar2016}
\bibinfo{author}{Boujibar, A.}, \bibinfo{author}{Bolfan-Casanova, N.},
  \bibinfo{author}{Andrault, D.}, \bibinfo{author}{Bouhifd, M.A.},
  \bibinfo{author}{Trcera, N.}, \bibinfo{year}{2016}.
\newblock \bibinfo{title}{{Incorporation of Fe$^{2+}$ and Fe$^{3+}$ in
  bridgmanite during magma ocean crystallization}}.
\newblock \bibinfo{journal}{American Mineralogist} \bibinfo{volume}{107},
  \bibinfo{pages}{1560--1570}.
\newblock \DOIprefix\doi{10.2138/am-2016-5561}.
%Type = Article
\bibitem[{Boukar\'e et~al.(2018)Boukar\'e, Parmentier and Parman}]{Boukare2018}
\bibinfo{author}{Boukar\'e, C.E.}, \bibinfo{author}{Parmentier, E.M.},
  \bibinfo{author}{Parman, S.W.}, \bibinfo{year}{2018}.
\newblock \bibinfo{title}{{Timing of mantle overturn during magma ocean
  solidification}}.
\newblock \bibinfo{journal}{Earth and Planetary Science Letters}
  \bibinfo{volume}{491}, \bibinfo{pages}{216--225}.
\newblock \DOIprefix\doi{10.1016/j.epsl.2018.03.037}.
%Type = Article
\bibitem[{Boukar\'e and Ricard(2017)}]{Boukare2017}
\bibinfo{author}{Boukar\'e, C.E.}, \bibinfo{author}{Ricard, Y.},
  \bibinfo{year}{2017}.
\newblock \bibinfo{title}{{Modeling phase separation and phase change for magma
  ocean solidification dynamics}}.
\newblock \bibinfo{journal}{Geochemistry, Geophysics, Geosystems}
  \bibinfo{volume}{18}, \bibinfo{pages}{3385--3404}.
\newblock \DOIprefix\doi{10.1002/2017GC006902}.
%Type = Article
\bibitem[{Boukar\'e et~al.(2015)Boukar\'e, Ricard and Fiquet}]{Boukare2015}
\bibinfo{author}{Boukar\'e, C.E.}, \bibinfo{author}{Ricard, Y.},
  \bibinfo{author}{Fiquet, G.}, \bibinfo{year}{2015}.
\newblock \bibinfo{title}{{Thermodynamics of the MgO-FeO-SiO$_{2}$ system up to
  140 GPa: Application to the crystallization of Earth's magma ocean}}.
\newblock \bibinfo{journal}{Journal of Geophysical Research: Solid Earth}
  \bibinfo{volume}{120}, \bibinfo{pages}{6085--6101}.
\newblock \DOIprefix\doi{10.1002/2015JB011929}.
%Type = Article
\bibitem[{Caracas et~al.(2019)Caracas, Hirose, Nomura and
  Ballmer}]{Caracas2019}
\bibinfo{author}{Caracas, R.}, \bibinfo{author}{Hirose, K.},
  \bibinfo{author}{Nomura, R.}, \bibinfo{author}{Ballmer, M.D.},
  \bibinfo{year}{2019}.
\newblock \bibinfo{title}{{Melt--crystal density crossover in a deep magma
  ocean}}.
\newblock \bibinfo{journal}{Earth and Planetary Science Letters}
  \bibinfo{volume}{516}, \bibinfo{pages}{202--211}.
\newblock \DOIprefix\doi{10.1016/j.epsl.2019.03.031}.
%Type = Article
\bibitem[{Catalli et~al.(2011)Catalli, Shim, Dera, Prakapenka, Zhao, Sturhahn,
  Chow, Xiao, Cynn and Evans}]{Catalli2011}
\bibinfo{author}{Catalli, K.}, \bibinfo{author}{Shim, S.H.},
  \bibinfo{author}{Dera, P.}, \bibinfo{author}{Prakapenka, V.B.},
  \bibinfo{author}{Zhao, J.}, \bibinfo{author}{Sturhahn, W.},
  \bibinfo{author}{Chow, P.}, \bibinfo{author}{Xiao, Y.},
  \bibinfo{author}{Cynn, H.}, \bibinfo{author}{Evans, W.J.},
  \bibinfo{year}{2011}.
\newblock \bibinfo{title}{{Effects of the Fe}$^{3+}${ spin transition on the
  properties of aluminous perovskite --- new insights for lower-mantle seismic
  heterogeneities}}.
\newblock \bibinfo{journal}{Earth and Planetary Science Letters}
  \bibinfo{volume}{310}, \bibinfo{pages}{293--302}.
\newblock \DOIprefix\doi{10.1016/j.epsl.2011.08.018}.
%Type = Article
\bibitem[{Catalli et~al.(2009)Catalli, Shim and Prakapenka}]{Catalli2009}
\bibinfo{author}{Catalli, K.}, \bibinfo{author}{Shim, S.H.},
  \bibinfo{author}{Prakapenka, V.B.}, \bibinfo{year}{2009}.
\newblock \bibinfo{title}{{Thickness and Clapeyron slope of the post-perovskite
  boundary}}.
\newblock \bibinfo{journal}{Nature} \bibinfo{volume}{462},
  \bibinfo{pages}{782--786}.
\newblock \DOIprefix\doi{10.1038/nature08598}.
%Type = Article
\bibitem[{Catalli et~al.(2010)Catalli, Shim, Prakapenka, Zhao, Sturhahn, Chow,
  Xiao, Liu, Cynn and Evans}]{Catalli2010}
\bibinfo{author}{Catalli, K.}, \bibinfo{author}{Shim, S.H.},
  \bibinfo{author}{Prakapenka, V.B.}, \bibinfo{author}{Zhao, J.},
  \bibinfo{author}{Sturhahn, W.}, \bibinfo{author}{Chow, P.},
  \bibinfo{author}{Xiao, Y.}, \bibinfo{author}{Liu, H.}, \bibinfo{author}{Cynn,
  H.}, \bibinfo{author}{Evans, W.J.}, \bibinfo{year}{2010}.
\newblock \bibinfo{title}{{Spin state of ferric iron in MgSiO}$_{3}${
  perovskite and its effect on elastic properties}}.
\newblock \bibinfo{journal}{Earth and Planetary Science Letters}
  \bibinfo{volume}{289}, \bibinfo{pages}{68--75}.
\newblock \DOIprefix\doi{10.1016/j.epsl.2009.10.029}.
%Type = Inbook
\bibitem[{Cobden et~al.(2015)Cobden, Thomas and Trampert}]{Cobden2015}
\bibinfo{author}{Cobden, L.}, \bibinfo{author}{Thomas, C.},
  \bibinfo{author}{Trampert, J.}, \bibinfo{year}{2015}.
\newblock \bibinfo{title}{Seismic Detection of Post-perovskite Inside the
  Earth}. \bibinfo{publisher}{Springer International Publishing}.
\newblock pp. \bibinfo{pages}{391--440}.
\newblock \DOIprefix\doi{10.1007/978-3-319-15627-9_13}.
%Type = Article
\bibitem[{Davies et~al.(2012)Davies, Goes, Davies, Schuberth, Bunge and
  Ristema}]{Davies2012}
\bibinfo{author}{Davies, D.R.}, \bibinfo{author}{Goes, S.},
  \bibinfo{author}{Davies, J.H.}, \bibinfo{author}{Schuberth, B.S.A.},
  \bibinfo{author}{Bunge, H.P.}, \bibinfo{author}{Ristema, J.},
  \bibinfo{year}{2012}.
\newblock \bibinfo{title}{{Reconciling dynamic and seismic models of Earth's
  lower mantle: The dominant role of thermal heterogeneity}}.
\newblock \bibinfo{journal}{Earth and Planetary Science Letters}
  \bibinfo{volume}{353--354}, \bibinfo{pages}{253--269}.
\newblock \DOIprefix\doi{10.1016/j.epsl.2012.08.016}.
%Type = Article
\bibitem[{Deschamps et~al.(2012)Deschamps, Cobden and Tackley}]{Deschamps2012b}
\bibinfo{author}{Deschamps, F.}, \bibinfo{author}{Cobden, L.},
  \bibinfo{author}{Tackley, P.J.}, \bibinfo{year}{2012}.
\newblock \bibinfo{title}{{The primitive nature of large low shear-wave
  velocity provinces}}.
\newblock \bibinfo{journal}{Earth and Planetary Science Letters}
  \bibinfo{volume}{349--350}, \bibinfo{pages}{198--208}.
\newblock \DOIprefix\doi{10.1016/j.epsl.2012.07.012}.
%Type = Article
\bibitem[{Deschamps et~al.(2019)Deschamps, Konishi, Fuji and
  Cobden}]{Deschamps2019}
\bibinfo{author}{Deschamps, F.}, \bibinfo{author}{Konishi, K.},
  \bibinfo{author}{Fuji, N.}, \bibinfo{author}{Cobden, L.},
  \bibinfo{year}{2019}.
\newblock \bibinfo{title}{{Radial thermo-chemical structure beneath Western and
  Northern Pacific from seismic waveform inversion}}.
\newblock \bibinfo{journal}{Earth and Planetary Science Letters}
  \bibinfo{volume}{520}, \bibinfo{pages}{153--163}.
\newblock \DOIprefix\doi{10.1016/j.epsl.2019.05.040}.
%Type = Article
\bibitem[{Deschamps and Tackley(2008)}]{Deschamps2008}
\bibinfo{author}{Deschamps, F.}, \bibinfo{author}{Tackley, P.J.},
  \bibinfo{year}{2008}.
\newblock \bibinfo{title}{{Searching for models of thermo-chemical convection
  that explain probabilistic tomography I. Principles and influence of
  rheological parameters.}}
\newblock \bibinfo{journal}{Physics of the Earth and Planetary Interiors}
  \bibinfo{volume}{171}, \bibinfo{pages}{357--373}.
\newblock \DOIprefix\doi{10.1016/j.pepi.2008.04.016}.
%Type = Article
\bibitem[{Elkins-Tanton et~al.(2003)Elkins-Tanton, Parmentier and
  Hess}]{ElkinsTanton2003}
\bibinfo{author}{Elkins-Tanton, L.T.}, \bibinfo{author}{Parmentier, E.M.},
  \bibinfo{author}{Hess, P.C.}, \bibinfo{year}{2003}.
\newblock \bibinfo{title}{{Magma ocean fractional crystallization and cumulate
  overturn in terrestrial planets: Implications for Mars}}.
\newblock \bibinfo{journal}{Meteoritics \& Planetary Science}
  \bibinfo{volume}{38}, \bibinfo{pages}{1753--1771}.
\newblock \DOIprefix\doi{10.1111/j.1945-5100.2003.tb00013.x}.
%Type = Article
\bibitem[{Fei et~al.(2007)Fei, Zhang, Corgne, Watson, Ricolleau, Meng and
  Prakapenka}]{Fei2007}
\bibinfo{author}{Fei, Y.W.}, \bibinfo{author}{Zhang, L.},
  \bibinfo{author}{Corgne, A.}, \bibinfo{author}{Watson, H.C.},
  \bibinfo{author}{Ricolleau, A.}, \bibinfo{author}{Meng, Y.},
  \bibinfo{author}{Prakapenka, V.B.}, \bibinfo{year}{2007}.
\newblock \bibinfo{title}{{Spin transition and equations of state of (Mg, Fe)O
  solid solutions}}.
\newblock \bibinfo{journal}{Geophysical Research Letters} \bibinfo{volume}{34},
  \bibinfo{pages}{L17307}.
\newblock \DOIprefix\doi{10.1029/2007GL030712}.
%Type = Article
\bibitem[{Gale et~al.(2013)Gale, Dalton, Langmuir, Su and Schilling}]{Gale2013}
\bibinfo{author}{Gale, A.}, \bibinfo{author}{Dalton, C.A.},
  \bibinfo{author}{Langmuir, C.H.}, \bibinfo{author}{Su, Y.},
  \bibinfo{author}{Schilling, J.G.}, \bibinfo{year}{2013}.
\newblock \bibinfo{title}{{The mean composition of ocean ridge basalts}}.
\newblock \bibinfo{journal}{Geochemistry, Geophysics, Geosystems}
  \bibinfo{volume}{14}, \bibinfo{pages}{489--518}.
\newblock \DOIprefix\doi{10.1029/2012GC004334}.
%Type = Article
\bibitem[{Garnero et~al.(2016)Garnero, McNamara and Shim}]{Garnero2016}
\bibinfo{author}{Garnero, E.J.}, \bibinfo{author}{McNamara, A.K.},
  \bibinfo{author}{Shim, S.H.}, \bibinfo{year}{2016}.
\newblock \bibinfo{title}{{Continent-sized anomalous zones with low seismic
  velocity at the base of Earth's mantle}}.
\newblock \bibinfo{journal}{Nature Geoscience} \bibinfo{volume}{9},
  \bibinfo{pages}{484--489}.
\newblock \DOIprefix\doi{10.1038/ngeo2733}.
%Type = Article
\bibitem[{Hirose et~al.(2017)Hirose, Morard, Sinmyo, Umemoto, Hernlund,
  Helffrich and Labrosse}]{Hirose2017}
\bibinfo{author}{Hirose, K.}, \bibinfo{author}{Morard, G.},
  \bibinfo{author}{Sinmyo, R.}, \bibinfo{author}{Umemoto, K.},
  \bibinfo{author}{Hernlund, J.}, \bibinfo{author}{Helffrich, G.},
  \bibinfo{author}{Labrosse, L.}, \bibinfo{year}{2017}.
\newblock \bibinfo{title}{{Crystallization of silicon dioxide and compositional
  evolution of the Earth's core}}.
\newblock \bibinfo{journal}{Nature} \bibinfo{volume}{543},
  \bibinfo{pages}{99--102}.
\newblock \DOIprefix\doi{10.1038/nature21367}.
%Type = Article
\bibitem[{Houser et~al.(2008)Houser, Masters, Shearer and Laske}]{Houser2008}
\bibinfo{author}{Houser, C.}, \bibinfo{author}{Masters, G.},
  \bibinfo{author}{Shearer, P.}, \bibinfo{author}{Laske, G.},
  \bibinfo{year}{2008}.
\newblock \bibinfo{title}{{Shear and compressional velocity models of the
  mantle from cluster analysis of long-period waveforms}}.
\newblock \bibinfo{journal}{Geophysical Journal International}
  \bibinfo{volume}{174}, \bibinfo{pages}{195--212}.
\newblock \DOIprefix\doi{10.1111/j.1365-246X.2008.03763.x}.
%Type = Article
\bibitem[{Ishii and Tromp(1999)}]{Ishii1999}
\bibinfo{author}{Ishii, M.}, \bibinfo{author}{Tromp, J.}, \bibinfo{year}{1999}.
\newblock \bibinfo{title}{{Normal-mode and free-air gravity constraints on
  lateral variations in velocity and density of Earth's mantle}}.
\newblock \bibinfo{journal}{Science} \bibinfo{volume}{285},
  \bibinfo{pages}{1231--1236}.
\newblock \DOIprefix\doi{10.1126/science.285.5431.1231}.
%Type = Article
\bibitem[{Jackson and Rigden(1996)}]{Jackson1996}
\bibinfo{author}{Jackson, I.}, \bibinfo{author}{Rigden, S.M.},
  \bibinfo{year}{1996}.
\newblock \bibinfo{title}{{Analysis of P-V-T data: constraints on the
  thermoelastic properties of high-pressure minerals}}.
\newblock \bibinfo{journal}{Physics of the Earth and Planetary Interiors}
  \bibinfo{volume}{96}, \bibinfo{pages}{85--112}.
\newblock \DOIprefix\doi{10.1016/0031-9201(96)03143-3}.
%Type = Article
\bibitem[{Koelemeijer et~al.(2016)Koelemeijer, Ritsema, Deuss and van
  Heijst}]{Koelemeijer2016}
\bibinfo{author}{Koelemeijer, P.}, \bibinfo{author}{Ritsema, J.},
  \bibinfo{author}{Deuss, A.}, \bibinfo{author}{van Heijst, H.J.},
  \bibinfo{year}{2016}.
\newblock \bibinfo{title}{{SP12RTS: a degree-12 model of shear- and
  compressional-wave velocity for Earth's mantle}}.
\newblock \bibinfo{journal}{Geophysical Journal International}
  \bibinfo{volume}{204}, \bibinfo{pages}{1024--1039}.
\newblock \DOIprefix\doi{10.1093/gji/ggv481}.
%Type = Article
\bibitem[{Labrosse et~al.(2007)Labrosse, Hernlund and Coltice}]{Labrosse2007}
\bibinfo{author}{Labrosse, S.}, \bibinfo{author}{Hernlund, J.W.},
  \bibinfo{author}{Coltice, N.}, \bibinfo{year}{2007}.
\newblock \bibinfo{title}{{A crystallizing dense magma ocean at the base of the
  Earth's mantle}}.
\newblock \bibinfo{journal}{Nature} \bibinfo{volume}{450},
  \bibinfo{pages}{866--869}.
\newblock \DOIprefix\doi{10.1038/nature06355}.
%Type = Article
\bibitem[{Lau et~al.(2017)Lau, Mitrovica, Davis, Tromp, Yang and
  Al-Attar}]{Lau2017}
\bibinfo{author}{Lau, H.C.P.}, \bibinfo{author}{Mitrovica, J.X.},
  \bibinfo{author}{Davis, J.L.}, \bibinfo{author}{Tromp, J.},
  \bibinfo{author}{Yang, H.Y.}, \bibinfo{author}{Al-Attar, D.},
  \bibinfo{year}{2017}.
\newblock \bibinfo{title}{{Tidal tomography constrains Earth's deep-mantle
  buoyancy}}.
\newblock \bibinfo{journal}{Nature} \bibinfo{volume}{551},
  \bibinfo{pages}{321--326}.
\newblock \DOIprefix\doi{10.1038/nature24452}.
%Type = Article
\bibitem[{Li et~al.(2006)Li, Weidner, Brodholt, Alf\`e, Price, Caracas and
  Wentzcovitch}]{Li2006c}
\bibinfo{author}{Li, L.}, \bibinfo{author}{Weidner, D.J.},
  \bibinfo{author}{Brodholt, J.P.}, \bibinfo{author}{Alf\`e, D.},
  \bibinfo{author}{Price, G.D.}, \bibinfo{author}{Caracas, R.},
  \bibinfo{author}{Wentzcovitch, R.}, \bibinfo{year}{2006}.
\newblock \bibinfo{title}{{Elasticity of CaSiO$_{3}$ perovskite at high
  pressure and high temperature}}.
\newblock \bibinfo{journal}{Physics of the Earth and Planetary Interiors}
  \bibinfo{volume}{155}, \bibinfo{pages}{249--259}.
\newblock \DOIprefix\doi{10.1016/j.pepi.2005.12.006}.
%Type = Article
\bibitem[{Li et~al.(2014)Li, Deschamps and Tackley}]{Li2014}
\bibinfo{author}{Li, Y.}, \bibinfo{author}{Deschamps, F.},
  \bibinfo{author}{Tackley, P.J.}, \bibinfo{year}{2014}.
\newblock \bibinfo{title}{{The stability and structure of primordial reservoirs
  in the lower mantle: insights from models of thermochemical convection in
  three-dimensional spherical geometry}}.
\newblock \bibinfo{journal}{Geophysical Journal International}
  \bibinfo{volume}{199}, \bibinfo{pages}{914--930}.
\newblock \DOIprefix\doi{10.1093/gji/ggu295}.
%Type = Inbook
\bibitem[{Masters et~al.(2000)Masters, Laske, Bolton and
  Dziewonski}]{Masters2000}
\bibinfo{author}{Masters, G.}, \bibinfo{author}{Laske, G.},
  \bibinfo{author}{Bolton, H.}, \bibinfo{author}{Dziewonski, A.},
  \bibinfo{year}{2000}.
\newblock \bibinfo{title}{{The relative behavior of shear velocity, bulk sound
  speed, and compressional velocity in the mantle: implications for chemical
  and thermal structure}}. \bibinfo{publisher}{AGU}. volume
  \bibinfo{volume}{117} of \textit{\bibinfo{series}{Geophysical Monograph
  Series}}.
\newblock pp. \bibinfo{pages}{63--87}.
\newblock \DOIprefix\doi{10.1029/GM117p0063}.
%Type = Article
\bibitem[{McDonough and Sun(1995)}]{McDonough1995}
\bibinfo{author}{McDonough, W.F.}, \bibinfo{author}{Sun, S.S.},
  \bibinfo{year}{1995}.
\newblock \bibinfo{title}{{The composition of the Earth}}.
\newblock \bibinfo{journal}{Chemical Geology} \bibinfo{volume}{120},
  \bibinfo{pages}{223--253}.
\newblock \DOIprefix\doi{10.1016/0009-2541(94)00140-4}.
%Type = Inbook
\bibitem[{McKay et~al.(1991)McKay, Heiken, Basu, Blanford, Simon, Reedy, French
  and Papike}]{McKay1991}
\bibinfo{author}{McKay, D.S.}, \bibinfo{author}{Heiken, G.},
  \bibinfo{author}{Basu, A.}, \bibinfo{author}{Blanford, G.},
  \bibinfo{author}{Simon, S.}, \bibinfo{author}{Reedy, R.},
  \bibinfo{author}{French, B.M.}, \bibinfo{author}{Papike, J.},
  \bibinfo{year}{1991}.
\newblock \bibinfo{title}{{The lunar regolith}}. \bibinfo{publisher}{Cambridge
  Univ. Press}.
\newblock pp. \bibinfo{pages}{285--356}.
%Type = Article
\bibitem[{McNamara and Zhong(2005)}]{McNamara2005}
\bibinfo{author}{McNamara, A.K.}, \bibinfo{author}{Zhong, S.},
  \bibinfo{year}{2005}.
\newblock \bibinfo{title}{{Thermochemical structures beneath Africa and the
  Pacific Ocean}}.
\newblock \bibinfo{journal}{Nature} \bibinfo{volume}{437},
  \bibinfo{pages}{1136--1139}.
\newblock \DOIprefix\doi{10.1038/nature04066}.
%Type = Article
\bibitem[{Murakami et~al.(2009)Murakami, Ohishi, Hirao and
  Hirose}]{Murakami2009}
\bibinfo{author}{Murakami, M.}, \bibinfo{author}{Ohishi, Y.},
  \bibinfo{author}{Hirao, N.}, \bibinfo{author}{Hirose, K.},
  \bibinfo{year}{2009}.
\newblock \bibinfo{title}{{Elasticity of MgO to 130~GPa: Implications for lower
  mantle mineralogy}}.
\newblock \bibinfo{journal}{Earth and Planetary Science Letters}
  \bibinfo{volume}{277}, \bibinfo{pages}{123--129}.
\newblock \DOIprefix\doi{10.1016/j.epsl.2008.10.010}.
%Type = Article
\bibitem[{Murakami et~al.(2012)Murakami, Ohishi, Hirao and
  Hirose}]{Murakami2012}
\bibinfo{author}{Murakami, M.}, \bibinfo{author}{Ohishi, Y.},
  \bibinfo{author}{Hirao, N.}, \bibinfo{author}{Hirose, K.},
  \bibinfo{year}{2012}.
\newblock \bibinfo{title}{{A perovskitic lower mantle inferred from
  high-pressure, high-temperature sound velocity data}}.
\newblock \bibinfo{journal}{Nature} \bibinfo{volume}{485},
  \bibinfo{pages}{90--94}.
\newblock \DOIprefix\doi{10.1038/nature11004}.
%Type = Article
\bibitem[{Nakagawa et~al.(2010)Nakagawa, Tackley, Deschamps and
  Connolly}]{Nakagawa2010b}
\bibinfo{author}{Nakagawa, T.}, \bibinfo{author}{Tackley, P.J.},
  \bibinfo{author}{Deschamps, F.}, \bibinfo{author}{Connolly, J.A.D.},
  \bibinfo{year}{2010}.
\newblock \bibinfo{title}{{The influence of MORB and harzburgite composition on
  thermo-chemical mantle convection in a 3-D spherical shell with
  self-consistently calculated mineral physics}}.
\newblock \bibinfo{journal}{Earth and Planetary Science Letters}
  \bibinfo{volume}{296}, \bibinfo{pages}{403--412}.
\newblock \DOIprefix\doi{10.1016/j.epsl.2010.05.026}.
%Type = Article
\bibitem[{Nomura et~al.(2011)Nomura, Ozawa, Tateno, Hirose, Hernlund, Muto,
  Ishii and Hiraoka}]{Nomura2011}
\bibinfo{author}{Nomura, R.}, \bibinfo{author}{Ozawa, H.},
  \bibinfo{author}{Tateno, S.}, \bibinfo{author}{Hirose, K.},
  \bibinfo{author}{Hernlund, J.}, \bibinfo{author}{Muto, S.},
  \bibinfo{author}{Ishii, H.}, \bibinfo{author}{Hiraoka, N.},
  \bibinfo{year}{2011}.
\newblock \bibinfo{title}{{Spin crossover and iron-rich silicate melt in the
  Earth's deep mantle}}.
\newblock \bibinfo{journal}{Nature} \bibinfo{volume}{473},
  \bibinfo{pages}{199--202}.
\newblock \DOIprefix\doi{10.1038/nature09940}.
%Type = Article
\bibitem[{Piet et~al.(2016)Piet, Badro, Nabiei, Dennenwaldt, Shim, Cantoni,
  H{\'e}bert and Gillet}]{Piet2016}
\bibinfo{author}{Piet, H.}, \bibinfo{author}{Badro, J.},
  \bibinfo{author}{Nabiei, F.}, \bibinfo{author}{Dennenwaldt, T.},
  \bibinfo{author}{Shim, S.H.}, \bibinfo{author}{Cantoni, M.},
  \bibinfo{author}{H{\'e}bert, C.}, \bibinfo{author}{Gillet, P.},
  \bibinfo{year}{2016}.
\newblock \bibinfo{title}{{Spin and valence dependence of iron partitioning in
  Earth's deep mantle}}.
\newblock \bibinfo{journal}{Proceedings of the National Academy of Sciences of
  the United States of America} \bibinfo{volume}{113},
  \bibinfo{pages}{11127--11130}.
\newblock \DOIprefix\doi{10.1073/pnas.1605290113}.
%Type = Article
\bibitem[{Ricolleau et~al.(2010)Ricolleau, Perillat, Fiquet, Daniel, Matas,
  Addad, Menguy, Cardon, Mezouar and Guignot}]{Ricolleau2010}
\bibinfo{author}{Ricolleau, A.}, \bibinfo{author}{Perillat, J.P.},
  \bibinfo{author}{Fiquet, G.}, \bibinfo{author}{Daniel, I.},
  \bibinfo{author}{Matas, J.}, \bibinfo{author}{Addad, A.},
  \bibinfo{author}{Menguy, N.}, \bibinfo{author}{Cardon, H.},
  \bibinfo{author}{Mezouar, M.}, \bibinfo{author}{Guignot, N.},
  \bibinfo{year}{2010}.
\newblock \bibinfo{title}{{Phase relations and equation of state of a natural
  MORB: Implications for the density profile of subducted oceanic crust in the
  Earth's lower mantle}}.
\newblock \bibinfo{journal}{Journal of Geophysical Research}
  \bibinfo{volume}{115}, \bibinfo{pages}{B08202}.
\newblock \DOIprefix\doi{10.1029/2009JB006709}.
%Type = Inbook
\bibitem[{Samuel et~al.(2005)Samuel, Farnetani and Andrault}]{Samuel2005}
\bibinfo{author}{Samuel, H.}, \bibinfo{author}{Farnetani, C.G.},
  \bibinfo{author}{Andrault, D.}, \bibinfo{year}{2005}.
\newblock \bibinfo{title}{{Heterogeneous lowermost mantle: compositional
  constraints and seismological observables}}. \bibinfo{publisher}{American
  Geophysical Union (AGU)}.
\newblock pp. \bibinfo{pages}{101--116}.
\newblock \DOIprefix\doi{10.1029/160GM08}.
%Type = Article
\bibitem[{Shim(2008)}]{Shim2008}
\bibinfo{author}{Shim, S.H.}, \bibinfo{year}{2008}.
\newblock \bibinfo{title}{{The Postperovskite Transition}}.
\newblock \bibinfo{journal}{Annual Review of Earth and Planetary Sciences}
  \bibinfo{volume}{36}, \bibinfo{pages}{569--599}.
\newblock \DOIprefix\doi{10.1146/annurev.earth.36.031207.124309}.
%Type = Article
\bibitem[{Shukla et~al.(2016)Shukla, Cococcioni and Wentzcovitch}]{Shukla2016}
\bibinfo{author}{Shukla, G.}, \bibinfo{author}{Cococcioni, M.},
  \bibinfo{author}{Wentzcovitch, R.M.}, \bibinfo{year}{2016}.
\newblock \bibinfo{title}{{Thermoelasticity of Fe$^{3+}$ - and Al-bearing
  bridgmanite: Effects of iron spin crossover}}.
\newblock \bibinfo{journal}{Geophysical Research Letters} \bibinfo{volume}{43},
  \bibinfo{pages}{5661--5670}.
\newblock \DOIprefix\doi{10.1002/2016GL069332}.
%Type = Article
\bibitem[{Shukla et~al.(2015)Shukla, Wu, Hsu, Floris, Cococcioni and
  Wentzcovitch}]{Shukla2015}
\bibinfo{author}{Shukla, G.}, \bibinfo{author}{Wu, Z.}, \bibinfo{author}{Hsu,
  H.}, \bibinfo{author}{Floris, A.}, \bibinfo{author}{Cococcioni, M.},
  \bibinfo{author}{Wentzcovitch, R.M.}, \bibinfo{year}{2015}.
\newblock \bibinfo{title}{{Thermoelasticity of Fe$^{2+}$-bearing bridgmanite}}.
\newblock \bibinfo{journal}{Geophysical Research Letters} \bibinfo{volume}{42},
  \bibinfo{pages}{1741--1749}.
\newblock \DOIprefix\doi{10.1002/2014GL062888}.
%Type = Inbook
\bibitem[{Solomatov(2015)}]{Solomatov2015}
\bibinfo{author}{Solomatov, V.}, \bibinfo{year}{2015}.
\newblock \bibinfo{title}{{Magma Oceans and Primordial Mantle
  Differentiation}}. \bibinfo{publisher}{Elsevier}. volume~\bibinfo{volume}{9}
  of \textit{\bibinfo{series}{Treatise on Geophysics}}.
\newblock pp. \bibinfo{pages}{81--104}.
\newblock \DOIprefix\doi{10.1016/B978-0-444-53802-4.00155-X}.
%Type = Article
\bibitem[{Su et~al.(1994)Su, Woodward and Dziewonski}]{Su1994}
\bibinfo{author}{Su, W.J.}, \bibinfo{author}{Woodward, R.L.},
  \bibinfo{author}{Dziewonski, A.M.}, \bibinfo{year}{1994}.
\newblock \bibinfo{title}{{Degree 12 model of shear velocity heterogeneity in
  the mantle}}.
\newblock \bibinfo{journal}{Journal of Geophysical Research}
  \bibinfo{volume}{99}, \bibinfo{pages}{6945--6980}.
\newblock \DOIprefix\doi{10.1029/93JB03408}.
%Type = Article
\bibitem[{Sun et~al.(2018)Sun, Wei, Han, Song, Li, Duan, Prakapenka and
  Mao}]{Sun2018}
\bibinfo{author}{Sun, N.}, \bibinfo{author}{Wei, W.}, \bibinfo{author}{Han,
  S.}, \bibinfo{author}{Song, J.}, \bibinfo{author}{Li, X.},
  \bibinfo{author}{Duan, Y.}, \bibinfo{author}{Prakapenka, V.B.},
  \bibinfo{author}{Mao, Z.}, \bibinfo{year}{2018}.
\newblock \bibinfo{title}{{Phase transition and thermal equations of state of
  (Fe, Al)-bridgmanite and post-perovskite: Implication for the chemical
  heterogeneity at the lowermost mantle}}.
\newblock \bibinfo{journal}{Earth and Planetary Science Letters}
  \bibinfo{volume}{490}, \bibinfo{pages}{161--169}.
\newblock \DOIprefix\doi{10.1016/j.epsl.2018.03.004}.
%Type = Incollection
\bibitem[{Tackley(1998)}]{Tackley1998b}
\bibinfo{author}{Tackley, P.J.}, \bibinfo{year}{1998}.
\newblock \bibinfo{title}{{Three-dimensional simulations of mantle convection
  with a thermochemical CMB boundary layer: D''?}}, in:
  \bibinfo{editor}{Gurnis, M.}, \bibinfo{editor}{Wysession, M.E.},
  \bibinfo{editor}{Knittle, E.}, \bibinfo{editor}{Buffett, B.A.} (Eds.),
  \bibinfo{booktitle}{{The Core-Mantle Boundary Region}}.
  \bibinfo{publisher}{AGU}, pp. \bibinfo{pages}{231--253}.
%Type = Article
\bibitem[{Tackley(2012)}]{Tackley2012}
\bibinfo{author}{Tackley, P.J.}, \bibinfo{year}{2012}.
\newblock \bibinfo{title}{{Dynamics and evolution of the deep mantle resulting
  from thermal, chemical, phase and melting effects}}.
\newblock \bibinfo{journal}{Earth-Science Reviews} \bibinfo{volume}{110},
  \bibinfo{pages}{1--25}.
\newblock \DOIprefix\doi{10.1016/j.earscirev.2011.10.001}.
%Type = Article
\bibitem[{Tateno and Hirose(2014)}]{Tateno2014}
\bibinfo{author}{Tateno, S.}, \bibinfo{author}{Hirose, K.~Ohishi, Y.},
  \bibinfo{year}{2014}.
\newblock \bibinfo{title}{{Melting experiments on peridotite to lowermost
  mantle conditions}}.
\newblock \bibinfo{journal}{Journal of Geophysical Research: Solid Earth}
  \bibinfo{volume}{119}, \bibinfo{pages}{4684--4694}.
\newblock \DOIprefix\doi{10.1002/2013JB010616}.
%Type = Article
\bibitem[{Thomson et~al.(2019)Thomson, Crichton, Brodholt, Wood, Siersch, Muir,
  Dobson and Hunt}]{Thomson2019}
\bibinfo{author}{Thomson, A.R.}, \bibinfo{author}{Crichton, W.A.},
  \bibinfo{author}{Brodholt, J.P.}, \bibinfo{author}{Wood, I.G.},
  \bibinfo{author}{Siersch, N.C.}, \bibinfo{author}{Muir, J.M.R.},
  \bibinfo{author}{Dobson, D.P.}, \bibinfo{author}{Hunt, S.A.},
  \bibinfo{year}{2019}.
\newblock \bibinfo{title}{{Seismic velocities of CaSiO$_{3}$ perovskite can
  explain LLSVPs in Earth's lower mantle}}.
\newblock \bibinfo{journal}{Nature} \bibinfo{volume}{572},
  \bibinfo{pages}{643--647}.
\newblock \DOIprefix\doi{10.1038/s41586-019-1483-x}.
%Type = Article
\bibitem[{Tolstikhin and Hofmann(2005)}]{Tolstikhin2005}
\bibinfo{author}{Tolstikhin, I.N.}, \bibinfo{author}{Hofmann, A.W.},
  \bibinfo{year}{2005}.
\newblock \bibinfo{title}{{Early crust on top of the Earth's core}}.
\newblock \bibinfo{journal}{Physics of the Earth and Planetary Interiors}
  \bibinfo{volume}{148}, \bibinfo{pages}{109--130}.
\newblock \DOIprefix\doi{10.1016/j.pepi.2004.05.011}.
%Type = Article
\bibitem[{Trampert et~al.(2004)Trampert, Deschamps, Resovsky and
  Yuen}]{Trampert2004}
\bibinfo{author}{Trampert, J.}, \bibinfo{author}{Deschamps, F.},
  \bibinfo{author}{Resovsky, J.}, \bibinfo{author}{Yuen, D.},
  \bibinfo{year}{2004}.
\newblock \bibinfo{title}{{Probabilistic Tomography Maps Chemical
  Heterogeneities Throughout the Lower Mantle}}.
\newblock \bibinfo{journal}{Science} \bibinfo{volume}{306},
  \bibinfo{pages}{853--856}.
\newblock \DOIprefix\doi{10.1126/science.1101996}.
%Type = Article
\bibitem[{Tronnes and Frost(2002)}]{Tronnes2002}
\bibinfo{author}{Tronnes, R.G.}, \bibinfo{author}{Frost, D.J.},
  \bibinfo{year}{2002}.
\newblock \bibinfo{title}{{Peridotite melting and mineral--melt partitioning of
  major and minor elements at 22--24.5 GPa}}.
\newblock \bibinfo{journal}{Earth and Planetary Science Letters}
  \bibinfo{volume}{197}, \bibinfo{pages}{117--131}.
\newblock \DOIprefix\doi{10.1016/S0012-821X(02)00466-1}.
%Type = Article
\bibitem[{Vilella(2020)}]{Vilella2020}
\bibinfo{author}{Vilella, K.}, \bibinfo{year}{2020}.
\newblock \bibinfo{title}{{Constraints on the composition and temperature of
  LLSVPs from seismic properties of lower mantle minerals}}.
\newblock \bibinfo{journal}{Mendeley Data} , \bibinfo{pages}{Version
  1}\DOIprefix\doi{10.17632/9w3n24sz6c.1}.
%Type = Article
\bibitem[{Vilella et~al.(2015)Vilella, Shim, Farnetani and
  Badro}]{Vilella2015b}
\bibinfo{author}{Vilella, K.}, \bibinfo{author}{Shim, S.H.},
  \bibinfo{author}{Farnetani, C.G.}, \bibinfo{author}{Badro, J.},
  \bibinfo{year}{2015}.
\newblock \bibinfo{title}{{Spin state transition and partitioning of iron:
  effects on mantle dynamics}}.
\newblock \bibinfo{journal}{Earth and Planetary Science Letters}
  \bibinfo{volume}{417}, \bibinfo{pages}{57--66}.
\newblock \DOIprefix\doi{10.1016/j.epsl.2015.02.009}.
%Type = Article
\bibitem[{Walter et~al.(2004)Walter, Nakamura, Tronnes and Frost}]{Walter2004}
\bibinfo{author}{Walter, M.J.}, \bibinfo{author}{Nakamura, E.},
  \bibinfo{author}{Tronnes, R.G.}, \bibinfo{author}{Frost, D.J.},
  \bibinfo{year}{2004}.
\newblock \bibinfo{title}{{Experimental constraints on crystallization
  differentiation in a deep magma ocean}}.
\newblock \bibinfo{journal}{Geochimica et Cosmochimica Acta}
  \bibinfo{volume}{68}, \bibinfo{pages}{4267--4284}.
\newblock \DOIprefix\doi{10.1016/j.gca.2004.03.014}.
%Type = Article
\bibitem[{Wang et~al.(2020)Wang, Xu, Sun, Ni, Wentzcovitch and Wu}]{Wang2020}
\bibinfo{author}{Wang, W.}, \bibinfo{author}{Xu, Y.}, \bibinfo{author}{Sun,
  D.}, \bibinfo{author}{Ni, S.}, \bibinfo{author}{Wentzcovitch, R.},
  \bibinfo{author}{Wu, Z.}, \bibinfo{year}{2020}.
\newblock \bibinfo{title}{{Velocity and density characteristics of subducted
  oceanic crust and the origin of lower-mantle heterogeneities}}.
\newblock \bibinfo{journal}{Nature Communications} \bibinfo{volume}{11},
  \bibinfo{pages}{64}.
\newblock \DOIprefix\doi{10.1038/s41467-019-13720-2}.
%Type = Article
\bibitem[{Wu et~al.(2013)Wu, Justo and Wentzcovitch}]{Wu2013}
\bibinfo{author}{Wu, Z.}, \bibinfo{author}{Justo, J.F.},
  \bibinfo{author}{Wentzcovitch, R.M.}, \bibinfo{year}{2013}.
\newblock \bibinfo{title}{{Elastic Anomalies in a Spin-Crossover System:
  Ferropericlase at Lower Mantle Conditions}}.
\newblock \bibinfo{journal}{Physical Review Letters} \bibinfo{volume}{110},
  \bibinfo{pages}{228501}.
\newblock \DOIprefix\doi{10.1103/PhysRevLett.110.228501}.

\end{thebibliography}
\end{document}